\documentclass[english,prd,floatfix,superscriptaddress,nofootinbib]{revtex4-2}
\usepackage[T1]{fontenc}
\usepackage[utf8]{inputenc}
\usepackage{color}
\definecolor{note_fontcolor}{rgb}{0.800781, 0.800781, 0.800781}
\usepackage{babel}
\usepackage{mathtools}
\usepackage{amsmath}
\usepackage{amssymb}
\usepackage{graphicx}
\usepackage{rotfloat}
\usepackage{booktabs}
\PassOptionsToPackage{hyphens}{url}\usepackage[unicode=true,pdfusetitle,
bookmarks=true,bookmarksnumbered=false,bookmarksopen=false,
breaklinks=false,pdfborder={0 0 0},pdfborderstyle={},backref=false,colorlinks=false]{hyperref}

\makeatletter

\usepackage{colortbl}
\usepackage{upgreek}
\usepackage{url}

\usepackage{pgfplots}
\usetikzlibrary{pgfplots.groupplots}

\DeclareSymbolFont{vectors}{OML}{cmm}{b}{it}
\DeclareSymbolFont{tensors}{OT1}{cmss}{bx}{it}

\DeclareSymbolFontAlphabet{\mathvec}{vectors}
\DeclareSymbolFontAlphabet{\mathtens}{tensors}
\DeclareUnicodeCharacter{03B2}{\ensuremath{\upbeta}}

\usepackage[capitalise]{cleveref}

\makeatother

\begin{document}
\global\long\def\tudu#1#2#3#4{{?{#1}^{#2}{}_{#3}{}^{#4}?}}%

\global\long\def\tud#1#2#3{{?{#1}^{#2}{}_{#3}?}}%

\global\long\def\tudud#1#2#3#4#5{{?{#1}^{#2}{}_{#3}{}^{#4}{}_{#5}?}}%

\global\long\def\tdu#1#2#3{\tensor{#1}{_{#2}^{#3}}}%

\global\long\def\dd#1#2{\frac{\mathrm{d}#1}{\mathrm{d}#2}}%

\global\long\def\pd#1#2{\frac{\partial#1}{\partial#2}}%

\global\long\def\tens#1{\mathtens{#1}}%

\global\long\def\threevec#1{\mathvec{#1}}%

\global\long\def\d{\mathrm{d}}%

\global\long\def\e{\mathrm{e}}%

\global\long\def\eps{\varepsilon}%

\global\long\def\i{\mathrm{i}}%

\global\long\def\ext{\tilde{\mathrm{d}}}%

\title{Breaking the north-south symmetry: \\ dyonic spinning black holes with synchronized gauged scalar hair}
\author{Pedro V. P. Cunha}
\email[]{pvcunha@ua.pt}
\affiliation{Departamento de Matemática da Universidade de Aveiro \\
and Centre for Research and Development in Mathematics and Applications (CIDMA) \\
Campus de Santiago, 3810-193 Aveiro, Portugal}
\author{Carlos A. R. Herdeiro}
\email[]{herdeiro@ua.pt}
\affiliation{Departamento de Matemática da Universidade de Aveiro \\
and Centre for Research and Development in Mathematics and Applications (CIDMA) \\
Campus de Santiago, 3810-193 Aveiro, Portugal}
\author{Eugen Radu}
\email[]{eugen.radu@ua.pt}
\affiliation{Departamento de Matemática da Universidade de Aveiro \\
and Centre for Research and Development in Mathematics and Applications (CIDMA) \\
Campus de Santiago, 3810-193 Aveiro, Portugal}
\author{Nuno M. Santos}
\email[]{nuno.m.santos@ua.pt}
\affiliation{Departamento de Matemática da Universidade de Aveiro \\
and Centre for Research and Development in Mathematics and Applications (CIDMA) \\
Campus de Santiago, 3810-193 Aveiro, Portugal}

\begin{abstract}
We study stationary clouds of a gauged, complex scalar field on a magnetically (and possibly electrically as well) charged Kerr-Newman black hole (BH). The existence of a magnetic charge $Q_m$ promotes a north-south \textit{asymmetry} of the scalar clouds. This breakdown of the clouds' $\mathbb{Z}_2$-symmetry carries through to the spacetime \textit{geometry} for the non-linear continuation of the clouds: a family of magnetically charged (or dyonic) BHs with synchronized gauged scalar hair, which we construct. Their distinct phenomenology is illustrated by their imaging, exhibiting skewed shadows and lensing.  Such hairy BHs could, in principle, result from the superradiant instability of magnetically charged Kerr-Newman BHs, unveiling a dynamical mechanism for creating north-south asymmetric BHs from standard $\mathbb{Z}_2$-symmetric electrovacuum BHs.
\end{abstract}
\maketitle

\tableofcontents
 
%\begin{abstract}
%
%\end{abstract}

%\linenumbers

\newpage 
%%%%%%%%%%%%%%%%%%%%%%%%%%%%%%%%%%%%%%%%%%%%%%%%%%%%%%%%%%%%%%%%%%%%%%%%%%%
% SECTION 
%%%%%%%%%%%%%%%%%%%%%%%%%%%%%%%%%%%%%%%%%%%%%%%%%%%%%%%%%%%%%%%%%%%%%%%%%%%
%%%%%%%%%%%%%%%%%%%%%%%%%%%%%%%%%%%%%%%%%%%%%%%%%%%%%%%%%%%%%%%%%%%%%%%%%%%
\section{Introduction}
%%%%%%%%%%%%%%%%%%%%%%%%%%%%%%%%%%%%%%%%%%%%%%%%%%%%%%%%%%%%%%%%%%%%%%%%%%%
%%%%%%%%%%%%%%%%%%%%%%%%%%%%%%%%%%%%%%%%%%%%%%%%%%%%%%%%%%%%%%%%%%%%%%%%%%%
%
The paradigmatic Kerr-Newman (KN) black holes (BHs) - see, e.g.,~\cite{Townsend:1997ku} - are north-south symmetric, with a well-defined equatorial plane as the set of fixed points of the $\mathbb{Z}_2$ north-south symmetry. Are there equilibrium (stationary), asymptotically flat BHs, with a connected horizon, which are \textit{not} $\mathbb{Z}_2$ north-south symmetric? And then, most importantly, is there any conceivable dynamical process that could form such exotic BHs?

There are indeed some known families of stationary, non-$\mathbb{Z}_2$ north-south symmetric BHs - see, e.g.,~\cite{Cunha:2018uzc,Cardoso:2018ptl,Chen:2020aix}. On the other hand, a recent proposal for using BH superradiance~\cite{Brito:2015oca} for the quest of magnetic monopoles~\cite{Pereniguez:2024fkn} opens a pathway to not only constructing non-$\mathbb{Z}_2$ north-south symmetric BHs but also a dynamical mechanism to form them. The purpose of this paper is to explore such non-standard BHs obtained in this context.

At the heart of the construction in this paper are magnetic monopoles, which are recurrent in theoretical high-energy physics and have a long and fascinating history - see, e.g.,~\cite{Shnir:2005vvi}. Some key developments that merit mentioning are the following. Dirac showed that the hypothetical existence of magnetic monopoles would explain electric charge quantization~\cite{Dirac:1931kp}. He derived that the magnetic charge $Q_m$ would related to the electron charge $q_e$ as (in Gauss units) 
\begin{equation}
    \frac{2q_e Q_m}{\hbar c}=N\in\mathbb{Z}\ .
    \label{Diracqc}
\end{equation}
This is known as the \textit{Dirac quantization condition}. It can also be derived by an alternative argument originally due to Saha - see, e.g.,~\cite{Balachandran:1993qz,Heras:2018uub}. Schwinger generalized this condition for particles with both electric and magnetic monopole charges, i.e., \textit{dyons}~\cite{Schwinger:1969ib}. 't Hooft~\cite{tHooft:1974kcl} and Polyakov~\cite{Polyakov:1974ek} later realized that magnetic monopoles are ubiquitous in Grand Unified Theories, where they can be non-singular solutions. Such theoretical developments led to magnetic monopoles (and also dyons) being routinely sought for, e.g., in cosmic rays and at colliders~\cite{ParticleDataGroup:2012pjm}. Despite extensive efforts, they have not yet been observed in nature, although they remain both a subject of theoretical interest and ongoing experimental searches. In a different effort, analogues of magnetic monopoles have been found in spin ices, realized as emergent, rather than elementary particles~\cite{Castelnovo:2007qi}.

In a recent development, it was argued that magnetic monopoles inside a rotating BH could act as natural amplifiers of their superradiant instability~\cite{Pereniguez:2024fkn}. Since the instability can dynamically lead to a new equilibrium BH state~\cite{East:2017ovw}, a BH with synchronized bosonic hair~\cite{Herdeiro:2017phl}, this suggests one should examine BHs with \textit{synchronized} hair and magnetic monopole charge. In this paper we start this analysis, first considering \textit{linear} (in the bosonic field, here taken to be a scalar field) equilibrium states, leading to the so-called stationary clouds~\cite{Hod:2012px}, and then their non-linear continuation~\cite{Herdeiro:2014goa}, which are BHs with synchronized gauged scalar hair and magnetic monopole charge - see~\cite{Delgado:2016jxq} for the case with electric charge only. 

The key observation of this paper is that the presence of a magnetic charge $Q_m$ leads to non-$\mathbb{Z}_2$ north-south symmetric scalar clouds on a dyonic KN BH. Then, as these clouds are made to backreact, the dyonic KN family (even with zero electric charge) bifurcates to a family of BHs with synchronized scalar hair which is non-$\mathbb{Z}_2$ north-south symmetric. To exemplify the impact of this spacetime feature on an observable, we study the gravitational lensing and shadows of the latter BHs, which are non-standard and skewed.

This paper is organized as follows. In Section~\ref{sec_model} we describe the action and field equations of the model to be studied. Section~\ref{section2} analyzes the linearized system (in the scalar field), taking a dyonic KN BH as the background. The corresponding stationary scalar clouds are obtained and explored. In Section~\ref{sec4} we construct the hairy BH solutions bifurcating from the KN solutions as the non-linear realization of the stationary clouds. Section~\ref{section5} presents some illustrative images of such BHs, i.e., their lensing and shadows. Finally, in Section~\ref{section5} we present our conclusions. Two appendices provide some technicalities on the monopole spherical harmonics and on the analytic computation of the scalar clouds for the extremal case. 

%\begin{itemize}
%    \item If they exist, magnetic monopoles should be created by the decay of sufficiently strong magnetic fields (by the ``dual" Schwinger mechanism).
%\end{itemize}

%%%%%%%%%%%%%%%%%%%%%%%%%%%%%%%%%%%%%%%%%%%%%%%%%%%%%%%%%%%%%%%%%%%%%%%%%%%%%%%
\section{The action and field equations}
\label{sec_model}
%%%%%%%%%%%%%%%%%%%%%%%%%%%%%%%%%%%%%%%%%%%%%%%%%%%%%%%%%%%%%%%%%%%%%%%%%%%%%%% 
 
%%%%%%%%%%%%%%%%%%%%%%%%%%%%%%%%%%%%%%%%%%%%%%%%%%%%%%%%%%%%%%%%%%%%%%%%%%%%%%%
%%%%%%%%%%%%%%%%%%%%%%%%%%%%%%%%%%%%%%%%%%%%%%%%%%%%%%%%%%%%%%%%%%%%%%%%%%%%%%% 

The action for Einstein--Maxwell theory minimally coupled to a massive complex scalar field $\Psi$ is 
\begin{align}
  \label{action}
  \mathcal{S} = \frac{1}{4 \pi}\int \text{d}^4x \sqrt{-g}
	\left[
	\frac{R}{4 G}
	-\frac{1}{4 }F_{ab}F^{ab}
	- {(D^a\Psi)^*}(D_a\Psi) - \mu^2\Psi\Psi^*   
	\right]\ , ~~{\rm with}~~ D_a=\nabla_a + i e A_a~,
\end{align}
where $R$ is the Ricci scalar of the metric $g_{ab}$, $g=\text{det}(g_{ab})$, $F_{ab}$ are the components of the Maxwell 2-form $F$, related to the 1-form potential $A=A_a\text{d}x^a$ as $F=\text{d}A$, and $D_a$ is the gauge covariant derivative. Additionally, $G$ is Newton's constant, $e$ is the gauge field coupling constant (i.e., the charge of the scalar field quanta), and $\mu$ is the scalar field mass. 

The variation of the action yields the Einstein--Maxwell--Klein-Gordon (EMKG) system,
\begin{align}
    G_{ab}  =2 GT_{ab}\ ,
    \quad
    D_{a}D^{a}\Psi=\mu^2 \Psi\ ,
    \quad
    \nabla_{a}F^{ab}=
i e \big [{(D^{b}\Psi)^*}\Psi-\Psi^*(D^b \Psi) \big ] 
\equiv e j^b  \ ,
\label{eom}
\end{align}
where $j^b$ is the scalar field current sourcing the Maxwell equations. The energy-momentum tensor includes independent contributions from the electromagnetic and scalar fields, 
$T=T_F+T_\Psi$, where the electromagnetic and scalar contributions respectively read
\begin{align}
    &(T_F)_{ab}\equiv  F_a^{~c}F_{bc} - \frac{1}{4}g_{ab}F_{cd}F^{cd} \ ,\\
    \label{Tab}
    &(T_\Psi)_{ab}\equiv {(D_{a}\Psi)^*} (D_{b}\Psi)
    +{(D_{b}\Psi)^*}(D_{a} \Psi) 
    -g_{ab}  \left\{ \frac{1}{2} g^{cd} 
    [{(D_{c}\Psi)^*}(D_{d} \Psi)+    
    {(D_{d}\Psi)^*}(D_{c} \Psi)]
    +\mu^2 \Psi^*\Psi\right\} \ .
\end{align}
The action in Eq.~\eqref{action} is invariant under a local $U(1)$ gauge transformation of the form
\begin{align}
    \Psi \to \Psi\,e^{-ie\alpha}\ ,
    \quad
    A_a\to A_a +\partial_a\alpha\ ,
    \label{eq:gauge}
\end{align}
where $\alpha$ is a real function of spacetime coordinates $x^a$.

%%%%%%%%%%%%%%%%%%%%%%%%%%%%%%%
%%%%%%%%%%%%%%%%%%%%%%%%%%%%%%%
%%%%%%%%%%%%%%%%%%%%%%%%%%%%%%%
\section{Linear solutions}
\label{section2}
%%%%%%%%%%%%%%%%%%%%%%%%%%%%%%%
%%%%%%%%%%%%%%%%%%%%%%%%%%%%%%%
%%%%%%%%%%%%%%%%%%%%%%%%%%%%%%%

Linearizing Eqs.~\eqref{eom} in $\Psi$, one obtains Einstein--Maxwell theory plus a decoupled Klein-Gordon equation. The dyonic KN BHs are solutions of the former. Stationary clouds are solutions of the latter when restricted to the dyonic KN BH background.

%%%%%%%%%%%%%%%%%%%%%%%%%%%%%%%
%%%%%%%%%%%%%%%%%%%%%%%%%%%%%%%
\subsection{Dyonic Kerr-Newman black holes\label{sec:1}}
%%%%%%%%%%%%%%%%%%%%%%%%%%%%%%%
%%%%%%%%%%%%%%%%%%%%%%%%%%%%%%%
In standard Boyer-Lindquist coordinates $(t,r,\theta,\varphi)$, the line element of the dyonic KN BH reads
\begin{align}
    \text{d}s^2=-\frac{\Delta}{\Sigma}\left(\text{d}t-a\sin^2\theta\,\text{d}\varphi\right)^2+\frac{\Sigma}{\Delta}\text{d}r^2+\Sigma\,\text{d}\theta^2+\frac{\sin^2\theta}{\Sigma}\left[a\,\text{d}t-\left(r^2+a^2\right)\text{d}\varphi\right]^2\ ,
    \label{dyonKN1}
\end{align}
where
\begin{align}
    \Sigma\equiv r^2+a^2\cos^2\theta\ ,
    \quad
    \Delta\equiv r^2-2Mr+a^2+Q_e^2+Q_m^2\ ,
\end{align}
while the Maxwell potential is
\begin{align}
    A_a\text{d}x^a=-\frac{Q_er}{\Sigma}\left(\text{d}t-a\sin^2\theta\,\text{d}\varphi\right)-\frac{Q_m\cos\theta}{\Sigma}\left[a\,\text{d}t-\left(r^2+a^2\right)\text{d}\varphi\right]\ .
     \label{dyonKN2}
\end{align}
Here, $a\equiv J/M$, where $M$ and $J$ are the Arnowitt–Deser–Misner (ADM) mass and angular momentum, respectively. Besides, $Q_e$ and $Q_m$ are the electric and magnetic charges of the BH. {Equation~\eqref{dyonKN2} corresponds to a choice of gauge such that the Maxwell potential vanishes at spatial infinity.}

The spacetime possesses two (commuting) Killing vectors, $\xi={\partial}_t$ and $\eta={\partial}_\varphi$, associated to stationarity and axisymmetry, respectively. The line element has coordinate singularities at $\Delta=0$ when $a^2+Q_e^2+Q_m^2\leq M^2$, which solves for $r=r_\pm=M\pm\sqrt{M^2-a^2-Q_e^2-Q_m^2}$. The hypersurface $r=r_+$ ($r=r_-$) is the outer (inner) Killing horizon. The angular velocity of the outer horizon is
\begin{align}
    \Omega_H=\frac{a}{r_+^2+a^2}\ .
\end{align}
The Killing vector $\xi$ is null on the ergosphere, i.e., on the hypersurface $r=r_\text{E}\equiv M+\sqrt{M^2-a^2\cos^2\theta-Q_e^2-Q_m^2}$. The ergosphere is timelike except where $\eta=0$, where it coincides with the outer horizon and becomes null. Also, $\xi$ is timelike outside the ergosphere and spacelike in the ergoregion, i.e., the spacetime region between the outer horizon and the ergosphere ($r_+<r<r_\text{E}$). Additionally, the axis of symmetry is defined by $\eta=0$.

The (co-rotating) electrostatic potential, $\Phi=-\chi^aA_a$, where $\chi^a=\xi^a+\Omega_H\eta^a$, is constant on the outer horizon, 
\begin{align}
    \Phi_H\equiv\left.\Phi\right|_{r=r_+}=\frac{Q_er_+}{r_+^2+a^2}\ .
\end{align}

In the absence of gravity ($M=a=0$) and electric charge ($Q_e=0$), the above solution reduces to Dirac's monopole, described by Minkowski metric and  
\begin{align}
    A_a\text{d}x^a=Q_m\cos\theta\,\text{d}\varphi \ ,
\end{align}
in spherical coordinates $(t,r,\theta,\varphi)$.

%%%%%%%%%%%%%%%%%%%%%%%%%%%%%%%%
%%%%%%%%%%%%%%%%%%%%%%%%%%%%%%%%%%%%%%%%%%%
\subsection{The decoupled Klein-Gordon equation}
%%%%%%%%%%%%%%%%%%%%
%%%%%%%%%%%%%%%%%%%%%%%%%%%%%%%%%%%%%%%%%%%%%%%%%%%%%%%

Explicitly, the Klein-Gordon equation in Eq.~\eqref{eom} reads
\begin{align}
    \Box\Psi+ie(\nabla_a A^a)\Psi+2ieA^a\nabla_a\Psi-e^2A_a A^a\Psi=\mu^2\Psi\ ,
\end{align}
where $\Box=\nabla^a\nabla_a$. One is interested in solving this equation on the background~\eqref{dyonKN1}--\eqref{dyonKN2}. The Maxwell potential in Eq.~\eqref{dyonKN2} satisfies the Lorenz condition, $\nabla_aA^a=0$. The existence of the Killing vectors $\{\xi,\eta\}$ guarantees the separation of the $t$-- and $\varphi$--dependencies in the scalar field. One can express the ansatz for modes solutions as
\begin{align}
    \Psi(t,r,\theta,\varphi)=e^{-i\omega t}\phi(r,\theta)e^{+im\varphi}\ ,
    \label{eq:scalar-ansatz}
\end{align}
{where $\omega>0$ is the mode frequency as measured by a static observer at spatial infinity, and $m$ is the azimuthal harmonic index, which can take either integer or half-integer values in this context (see discussion below).} 

%\begin{align}
%    &\Box\Phi=\left(-g^{tt}\omega^2-g^{\varphi\varphi}m^2+2g^{t\varphi}m\omega\right)\Phi+\frac{1}{\Sigma}\partial_t\left(\Sigma\,g^{rr}\partial_r\Phi\right)+\frac{1}{\Sigma\sin\theta}\partial_\theta\left(\Sigma\sin\theta\,g^{\theta\theta}\partial_\theta \Phi\right)\ ,\\
%    &ie(\nabla_\mu A^\mu)\Phi=0\ ,\\
%    &2ieA^\mu\nabla_\mu\Phi=\frac{1}{\Sigma}\left[\frac{2eQr[(r^2+a^2)\omega-ma]}{\Delta}+2eP(m-a\omega\sin\theta)\frac{\cos\theta}{\sin^2\theta}\right]\Phi\ ,\\
%    &-e^2A_\mu A^\mu\Phi=-\frac{e^2}{\Sigma}\left(-\frac{Q^2r^2}{\Delta}+P^2\cot^2\theta\right)\Phi=-\frac{e^2}{\Sigma}\left(-\frac{Q^2r^2}{\Delta}-P^2+\frac{P^2}{\sin^2\theta}\right)\Phi
%\end{align}

Plugging the ansatz into the Klein-Gordon equation restricted to the background \eqref{dyonKN1}--\eqref{dyonKN2}, one obtains
\begin{align}
    &\left[\frac{(r^2+a^2)^2}{\Delta}-a^2\sin^2\theta\right]\omega^2+\left[\frac{a^2}{\Delta}-\frac{1}{\sin^2\theta}\right]m^2+2\left[1-\frac{r^2+a^2}{\Delta}\right]ma\omega+\frac{1}{\phi}\partial_r\left(\Delta\,\partial_r\phi\right)+\nonumber\\
    &+\frac{1}{\phi\sin\theta}\partial_\theta\left(\sin\theta\,\partial_\theta \phi\right)+\frac{2eQ_er}{\Delta}\left[(r^2+a^2)\omega-ma\right]-2eQ_m(m-a\omega\sin^2\theta)\frac{\cos\theta}{\sin^2\theta}+\frac{e^2Q_e^2r^2}{\Delta}+\nonumber\\
    &+e^2Q_m^2-\frac{e^2Q_m^2}{\sin^2\theta}=\mu^2(r^2+a^2\cos^2\theta)\ .
\end{align}
Inspection of this equation reveals that the $r$-- and $\theta$--dependencies can be separated, which means one can take 
\begin{align}
    \phi(r,\theta)=R(r)S(\theta)\ .
\end{align}
Writing $a^2\omega^2\sin^2\theta=a^2\omega^2(1-\cos^2\theta)$, keeping all $\theta$--dependent terms in the left-hand side and moving all $r$--dependent and constant terms to the right-hand side, one gets
\begin{align}
    &\frac{1}{S(\theta)\sin\theta}\frac{\text{d}}{\text{d}\theta}\left[\sin\theta\frac{\text{d}S(\theta)}{\text{d}\theta}\right]+(\omega^2-\mu^2)a^2\cos^2\theta-\frac{m^2+2eQ_m(m-a\omega\sin^2\theta)\cos\theta+e^2Q_m^2}{\sin^2\theta}=\nonumber\\
    &=-\frac{1}{R}\frac{\text{d}}{\text{d}r}\left(\Delta\frac{\text{d}R(r)}{\text{d}r}\right)-\frac{\left[\left(r^2+a^2\right)\omega-ma+eQ_er\right]^2}{\Delta}+a^2\omega^2-2ma\omega+\mu^2r^2-e^2Q_m^2 \ .
\end{align}
Each side of the equation must be equal to a constant, here denoted by $-\Lambda$. One then gets two second-order ordinary differential equations~\cite{Semiz:1991kh},
\begin{align}
    &\frac{1}{\sin\theta}\frac{\text{d}}{\text{d}\theta}\left[\sin\theta\frac{\text{d}S(\theta)}{\text{d}\theta}\right]+\left[a^2\omega^2-2ma\omega-a^2\mu^2\cos^2\theta-\frac{(m+eQ_m\cos\theta-a\omega\sin^2\theta)^2}{\sin^2\theta}-e^2Q_m^2+\Lambda\right]S(\theta)=0\ ,\\
    &\frac{\text{d}}{\text{d}r}\left(\Delta\frac{\text{d}R(r)}{\text{d}r}\right)+\left\{\frac{\left[(r^2+a^2)\omega-ma+eQ_er\right]^2}{\Delta}-a^2\omega^2+2ma\omega-\mu^2r^2+e^2Q_m^2-\Lambda\right\}R(r)=0\ .
\end{align}
These equations are both confluent Heun equations: the former (latter) has singular points at $r=r_\pm$ ($\theta=0,\pi$). They are coupled via the Killing eigenvalues $\{\omega,m\}$, the background parameters $\{a,Q_e,Q_m\}$ and the separation constant $\Lambda$. The system remains invariant under the discrete transformation $\{a\omega,m,Q_e,Q_m\}\rightarrow\{-a\omega,-m,-Q_e,-Q_m\}$. 

When $a=0$, the angular equation reduces to the Jacobi differential equation and the angular dependence of the scalar field is described by the \textit{monopole spherical harmonics} $Y_{q,\ell,m}$ (see Appendix~\ref{App:1}), provided that $\Lambda=\ell(\ell+1)$, with $\ell=|q|,|q|+1,\ldots$ and $m=-\ell,-\ell+1,\ldots,\ell-1,\ell$, where
\begin{align} 
q\equiv e Q_m=\frac{N}{2} \ , \qquad {N\in \mathbb{Z}} \ ,
\end{align} 
where the last equality is precisely the Dirac quantization condition in Eq.~\eqref{Diracqc}. Since $e$ is interpreted as the charge of the scalar field quanta, {$|N|$ is to be interpreted as the \textit{number of magnetic monopoles}}. This relation shows that $q$ is an integer or half-integer, and so are $\ell$ and $m$. 

For $a\neq0$, the angular eigenfunctions are the \textit{monopole spheroidal harmonics}~\cite{Semiz:1991kh}, which depend on $\{q,\ell,m\}$, on the product $a\omega$ and on the \textit{spheroidicity} 
\begin{equation}
c\equiv |a|\sqrt{\omega^2-\mu^2} \ ,
\end{equation} 
i.e., $S=S_{q,\ell,m}(a\omega,c,\theta)$. Likewise, $\Lambda=\Lambda_{q,\ell,m}(a\omega,c)$. {As opposed to the conventional spheroidal harmonics, their monopole counterparts do not exhibit a $\mathbb{Z}_2$ north-south symmetry (see details in Appendix~\ref{App:1}).}

Concerning the radial equation, the radial function behaves as
\begin{align}
    \left.R\right|_{y\rightarrow-\infty}\sim e^{\pm i\omega_\star y}\ ,
    \quad
    \left.R\right|_{y\rightarrow+\infty}\sim y^{-1}e^{\pm\sqrt{\mu^2-\omega^2}y}\ ,
    \label{eq:asympt}
\end{align}
where $\omega_\star\equiv \omega-\omega_c$, with $\omega_c\equiv m\Omega_H-e\Phi_H$, and $y$ is the tortoise coordinate, defined by
\begin{align}
    y(r)=r+\frac{r_+^2}{r_+-r_-}\log(r-r_+)-\frac{r_-^2}{r_+-r_-}\log(r-r_-)\ ,
\end{align}
which approaches $-\infty$ ($+\infty$) as $r$ approaches $r_+$ ($+\infty$). One is interested in mode solutions of the Klein-Gordon equation that ($i$) satisfy the \textit{synchronization condition},
\begin{align}
    \omega=\omega_c \ ,
    \label{sync} 
\end{align}
and ($ii$) vanish at spatial infinity. Condition ($i$) is equivalent to requiring the radial function to be regular at $r=r_+$, i.e., to have a Taylor series around $r=r_+$,
\begin{align}
    \left.R\right|_{r\rightarrow r_+} \sim \sum_{j=0}^\infty c_{(j)}(r-r_+)^j\ ,
\end{align}
where, without loss of generality, $c_{(0)}=1$ and the coefficients $\{c_{(j)}\}_{j\in\mathbb{N}_0}$ are obtained by solving the radial equation order by order. On the other hand, condition ($ii$) amounts to imposing\footnote{The spheroidicity is purely imaginary in this case.}
\begin{equation}
\omega^2\leq \mu^2 \ ,    
\label{bound}
\end{equation} 
so that the radial function decays exponentially in space. This is different from Eq.~\eqref{condw} below because of a different gauge for $A$, as explained in Section~\ref{bond}. Solutions obeying conditions ($i$)--($ii$) are bound states - here dubbed \textit{stationary scalar clouds}.

\subsection{Stationary scalar clouds}
The dyonic KN BHs are described by four parameters, the four global charges $\{M,J,Q_e,Q_m\}$. When studying them, one typically takes the mass $M$ as a scale yielding an effective 3-dimensional parameter space. On the other hand, when considering the massive Klein-Gordon equation on this background, the scalar field mass $\mu$ provides a more natural overall scale, as this is fixed by the theory rather, and, unlike $M$, is not an integration constant. It is convenient, then, to analyze the equilibrium bound solutions between the scalar field and dyonic KN BHs (i.e., stationary scalar clouds) in terms of the \textit{dimensionless} four-parameter space $\{M\mu,J\mu^2,Q_e\mu,Q_m\mu\}$. 
%Henceforth, when referring to the background parameters, we shall mean these dimensionless ones.

Stationary clouds are only supported in a subset of the above parameter space. Fixing the background gauge charges, $\{Q_e\mu,Q_m\mu\}$, as well as the gauge coupling and the angular ``quantum'' numbers, $\{q,\ell,m\}$, bound states only exist along lines in the 2-dimensional parameter space spanned by $\{M\mu,J\mu^2\}$ or, equivalently, $\{M\mu,\Omega_H/\mu\}$. These are known as \textit{existence lines}~\cite{Herdeiro:2014goa}, that terminate in extremal BHs, the endpoint(s) being known as \textit{Hod point(s)}~\cite{Herdeiro:2015tia}. Different existence lines are labeled by the number of nodes in the radial direction, $n$, i.e., the number of nodes of the radial function. 

The following analysis concerns the electrically neutral case only ($Q_e=0$). When $\omega=\omega_c$, the radial and angular equations remain invariant under the transformation $\{m,\theta\}\rightarrow\{-m,\pi-\theta\}$, which means that $\Lambda_{q,\ell,m}(a\omega_c,c)=\Lambda_{q,\ell,-m}(a\omega_c,c)$ and, consequently, the radial function remains invariant and $S_{n,q,\ell,m}(a\omega_c,c,\theta) =S_{n,q,\ell,-m}(a\omega_c,c,\pi-\theta)$. In other words, one can get the bound state $\{n,q,\ell,-m\}$ by mirroring the bound state $\{n,q,\ell,m\}$ with respect to the equatorial plane, $\theta=\pi/2$. {Thus, the results to be presented are restricted to $m>0$.} 

\begin{figure}[h!]
    \centering
    \includegraphics[scale=0.9]{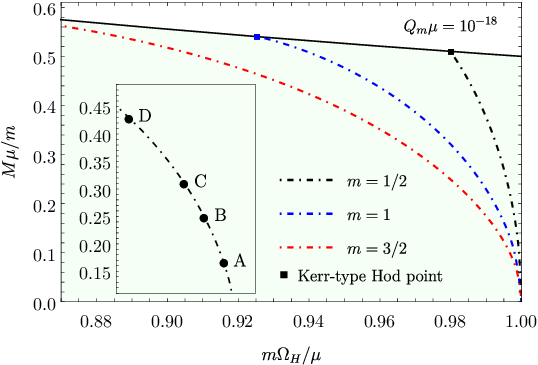}
    \caption{Existence lines for $Q_e\mu=0$, $Q_m\mu=10^{-18}$, $n=0$ and $|q|=\ell=m=1/2,1,3/2$. The squares are Kerr-type Hod points. The circles in the inset, dubbed A--D, are illustrative solutions with $m=1/2$, listed in Table~\ref{tab:1} and represented in Fig.~\ref{fig:2}.}
    \label{fig:1}
\end{figure}

\begin{figure}[ht]
    \centering
    \includegraphics[scale=0.9]{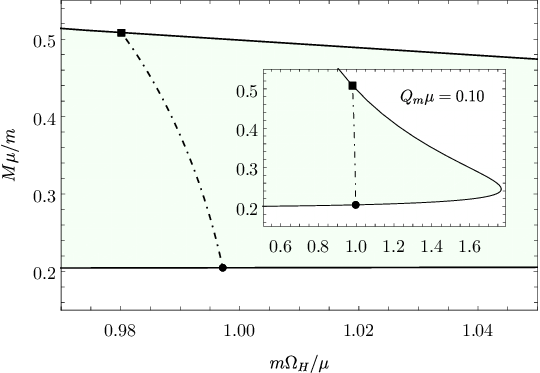}
    \qquad
    \includegraphics[scale=0.9]{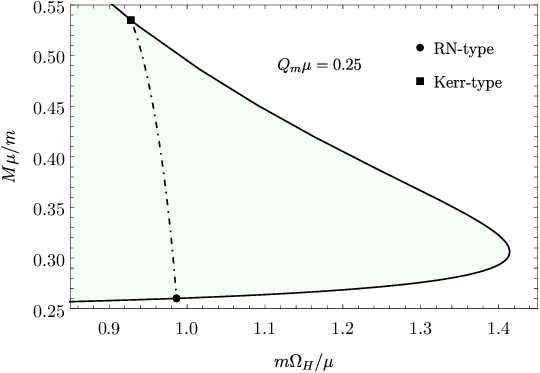}
    \caption{(Left) Existence line for $Q_e\mu=0$, $Q_m\mu=0.10$, $n=0$ and $|q|=\ell=m=1/2$. (Right) Existence line for $Q_e\mu=0$, $Q_m\mu=0.25$, $n=0$ and $|q|=\ell=m=1$. The circles (squares) are Reissner-Nordstr\"om-(Kerr-)type Hod points.}
    \label{fig:moderate-P}
\end{figure}

The existence lines for\footnote{This is the value of $Q_m\mu$ used in~\cite{Pereniguez:2024fkn} for charged pions.} $Q_m\mu=10^{-18}$ and $|q|=\ell=m=1/2,1,3/2$ are plotted in Fig.~\ref{fig:1}. They correspond respectively to $|N|=1,2,3$ magnetic monopoles in the ground state ($n=0$). The light green shaded region represents the parameter space of dyonic KN BHs with $Q_e\mu=0$ and $Q_m\mu=10^{-18}$, and the black solid line refers to extremal BHs ($M^2=a^2+Q_m^2$). The existence line joins two Hod points, each corresponding to a scalar cloud around an extremal dyonic BH.\footnote{Purely electric KN BHs also have two Hod points. In~\cite{Benone:2014ssa}, it is said that ``a distinct feature of the charged existence lines [\ldots] is that the existence lines do not reach $M=0$, since the inclusion of background charge implies a minimum value for the background mass, i.e. $|Q|<M$''. This minimum corresponds to an extremal KN BH, leading to the second Hod point along existence lines, albeit not explicitly discussed in~\cite{Benone:2014ssa}.}
{The Hod point with the greatest (lowest) value of $a\mu$ is said to be of Kerr-(Reissner-Nordstr\"om-)type.} These can be found analytically (see~Appendix \ref{App:2}). The existence of two Hod points is not clear in Fig. \ref{fig:1}, as $Q_m\mu$ is tiny, but becomes apparent for moderate values of $Q_m\mu$ - see Fig.~\ref{fig:moderate-P}. 


Figure \ref{fig:1} reveals that the existence lines move to the left as $m$ increases, in agreement with~\cite{Herdeiro:2014goa}. On the other hand, fixing $m$, they move to the right as either $n$ and/or $\ell$ increases (not shown here).

\begin{figure}[h!]
    \centering
    \begin{minipage}[b]{.12\linewidth}
        \centering
        \includegraphics[width=\columnwidth]{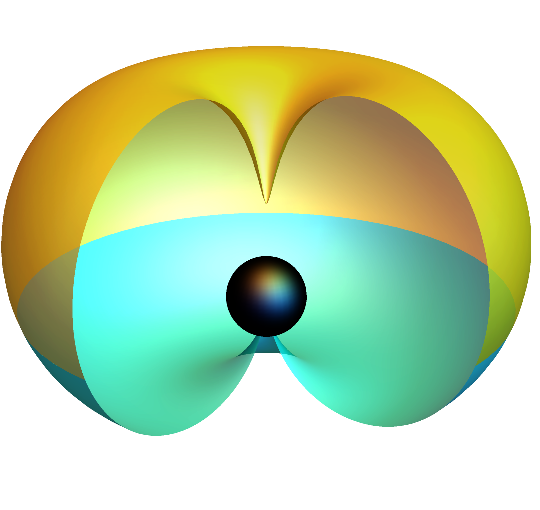}
        A
    \end{minipage}
    \qquad
    \begin{minipage}[b]{.12\linewidth}
        \centering
        \includegraphics[width=\columnwidth]{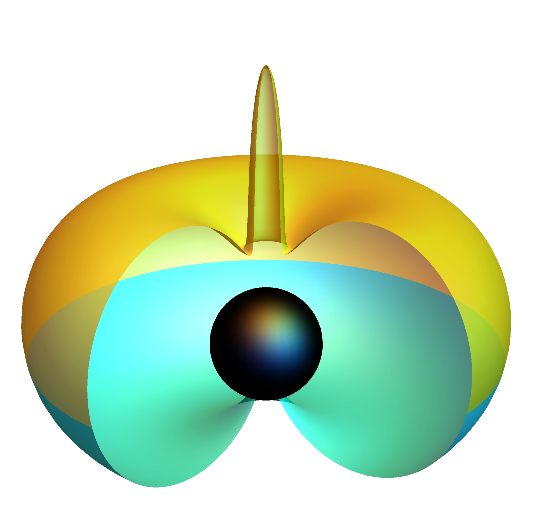}
        B
    \end{minipage}
    \qquad
    \begin{minipage}[b]{.12\linewidth}
        \centering
        \includegraphics[width=\columnwidth]{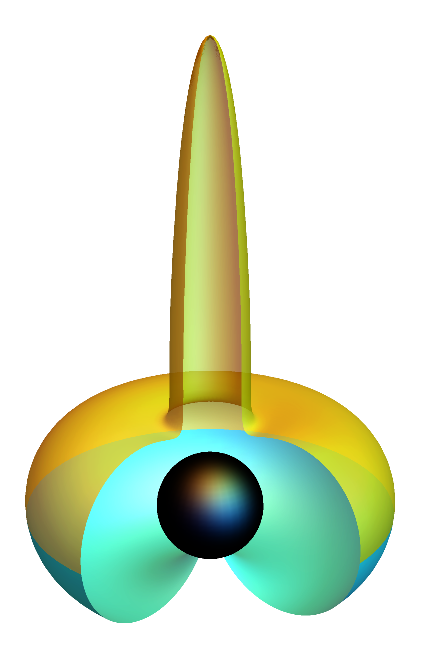}
        C
    \end{minipage}
    \qquad
    \begin{minipage}[b]{.12\linewidth}
        \centering
        \includegraphics[width=\columnwidth]{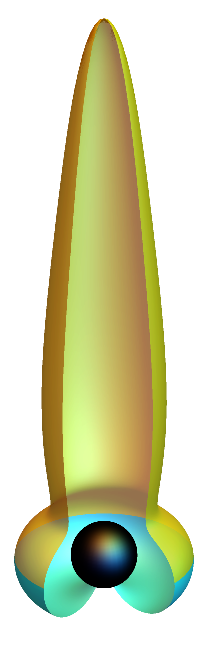}
        D
    \end{minipage}
    \caption{Three-dimensional plots of spherical radius $|S(\theta)e^{+im\varphi}|$ for the illustrative clouds in Fig. \ref{fig:1} and Table~\ref{tab:1}, where $\theta\in[0,\pi]$ and $\varphi\in[0,2\pi)$ are taken to be the usual spherical coordinates. For ease of visualization, $\varphi$ varies from $0$ to $3\pi/2$. Additionally, the two hemispheres are shaded in different colors.
    %\Santos{NS: The solutions Eugen found so far (mostly for $\omega/\mu=0.99$) are compatible with A. I asked him to probe smaller $\omega$ so that we can check the morphology there matches what I found.}
    }
    \label{fig:2}
\end{figure}

The properties of the illustrative scalar cloud solutions A--D in Fig.~\ref{fig:1} are listed in Table \ref{tab:1}. Their spatial distribution, shown in Fig. \ref{fig:2}, unequivocally reveals the breaking of the north-south symmetry. From a technical viewpoint, this can be understood by noticing that, for $m\neq 0$ and $Q_m\neq 0$, the component $D_\varphi \Psi = i(m+e A_\varphi) \Psi$ of the gauge covariant derivative {\it cannot be} invariant under the transformation $\theta \to \pi-\theta$, since the term $(m+eA_\varphi)$ is (asymptotically) the linear combination of a $\mathbb{Z}_2$-even and a $\mathbb{Z}_2$-odd term.
Then, the (gauged) scalar field energy-momentum tensor is not $\mathbb{Z}_2$-symmetric either, which will be relevant in the next section.
As seen in Fig. \ref{fig:2}, as $\omega/\mu$ decreases, $a\mu$ increases, resulting in the spread of the scalar field along the axis of symmetry in one of the hemispheres. 
\begin{table}[h]
    \centering
    \begin{tabular}{c|ccccc}
        \toprule
        \toprule
        Solution & $r_+\mu$ & $M\mu$ & $a\mu$ & $\omega/\mu$ & $\Lambda$ \\
        \hline
        A & $0.150$ & $0.0833$ & $0.0499$ & $0.9981$ & $0.7664$ \\
        B & $0.200$ & $0.1246$ & $0.0993$ & $0.9958$ & $0.7823$ \\
        C & $0.225$ & $0.1554$ & $0.1389$ & $0.9933$ & $0.7946$ \\
        D & $0.250$ & $0.2149$ & $0.2120$ & $0.9866$ & $0.8168$ \\
        \bottomrule
        \bottomrule
    \end{tabular}
    \caption{Properties of the solutions A--D identified in Fig.~\ref{fig:1}.}
    \label{tab:1}
\end{table}

\begin{figure}[ht]
    \centering
    \includegraphics[scale=0.9]{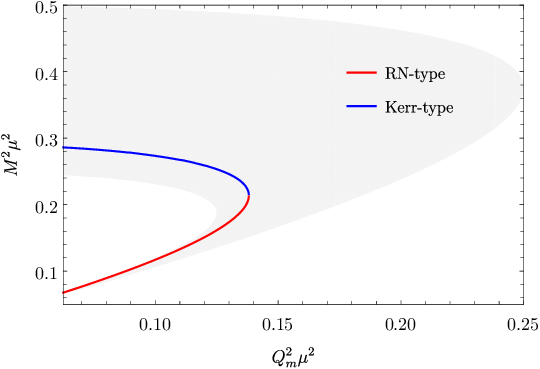}
    \caption{Hod points in a mass-magnetic charge diagram for extremal purely magnetic KN BHs ($Q_e=0$, $M^2=a^2+Q_m^2$) with $\ell=m=1$. Reissner-Nordstr\"om-(Kerr-)type Hod points are shown in red (blue). They meet at $Q_m\mu\approx 0.3715$. The gray shaded region is where conditions ($i$) and ($ii$) are satisfied (see main text and Appendix~\ref{App:2}).}
    \label{fig:extremal}
\end{figure}

The existence line shrinks as $Q_m\mu$ increases and becomes a point for a critical value of $Q_m\mu$, above which no clouds are found. For $\ell=m=1$, the critical value is $Q_m\mu\approx0.3715$ -- see Fig. \ref{fig:extremal}.

        % \begin{align}
        %     \sqrt{\frac{m^2}{4} + \frac{Q_m^2\mu^2}{2} - \frac{|m|}{4}\sqrt{m^2-4Q_m^2\mu^2}} < M\mu < \sqrt{\frac{m^2}{4} + \frac{Q_m^2\mu^2}{2} + \frac{|m|}{4}\sqrt{m^2-4Q_m^2\mu^2}}\ ,
        % \end{align}
        % where
        % \begin{align}
        %     |Q_m\mu|<\frac{|m|}{2}\ .
        % \end{align}
        % On the other hand, $\epsilon^2 \geq 0$, which implies
        %  \begin{align}
        %     \label{cond:2}
        %     M\mu \leq \sqrt{\frac{m^2}{8} + \frac{Q_m^2\mu^2}{2} - \frac{|m|}{8}\sqrt{m^2-8Q_m^2\mu^2}}
        %     \quad
        %     \lor
        %     \quad
        %     M\mu \geq \sqrt{\frac{m^2}{8} + \frac{Q_m^2\mu^2}{2} + \frac{|m|}{8}\sqrt{m^2-8Q_m^2\mu^2}}\ ,
        % \end{align}   
        % %
        % with
        % \begin{align}
        %     \label{cond:3}
        %     |Q_m\mu| \leq \frac{|m|}{2\sqrt{2}}\ .
        % \end{align}
        % For $m=\frac{1}{2}$, $|Q_m\mu| \leq \frac{1}{4\sqrt{2}}\approx 0.177$. For $m=1$, $|Q_m\mu| \leq \frac{1}{2\sqrt{2}}\approx 0.354$.
        
        % If $Q_m=0$, one gets
        % \begin{align}
        %     \frac{|m|}{2}<M\mu<\frac{|m|}{\sqrt{2}}\ . 
        % \end{align}
        % For non-vanishing $P$, one can find solutions in both regions defined by Eq.~\eqref{cond:2}: the two Hod points. There should be solutions saturating Eq.~\eqref{cond:3}.

\section{Non-linear solutions}
\label{sec4}
%%%%%%%%%%%%%%%%%%%%%%%%%%%%%%%%%%%%%%%%%%%%%%%%%%%
%%%%%%%%%%%%%%%%%%%%%%%%%%%%%%%%%%%%%%%%%%%%%%%%%%%

The zero modes are a linear manifestation that the dyonic KN BHs bifurcate to a new family of fully non-linear stationary solutions with scalar hair~\cite{Herdeiro:2014ima,Dias:2015nua}. Moreover, from the previous analysis, it becomes clear that the backreaction of the stationary clouds will break the north-south $\mathbb{Z}_2$ symmetry of the geometry in such new family of BHs.

%%%%%%%%%%%%%%%%%%%%%%%%%%%%%%%%%%%%%%%%%%%%%%%%%%%%%%%%%%%%%%%%%%%%%%%%%%%%%%%
\subsection{Ans\"atze} 
%%%%%%%%%%%%%%%%%%%%%%%%%%%%%%%%%%%%%%%%%%%%%%%%%%%%%%%%%%%%%%%%%%%%%%%%%%%%%%% 
%
One seeks hairy BH solutions that are regular on and outside the event horizon, stationary, axisymmetric, and asymptotically flat. These possess two (commuting) Killing vectors, {$\{\xi,\eta\}$}, associated to stationarity and axisymmetry, respectively. Thus, coordinates $(t,\varphi)$ adapted to the Killing vectors {$\xi=\partial_t$} and {$\eta=\partial_\varphi$} can be chosen. It is also assumed that there exists a two-space orthogonal to $\{\xi,\eta\}$, in which spherical-like coordinates $(r,\theta)$ can be introduced so that $r$ is orthogonal to $\theta$. A line element compatible with these assumptions is
\begin{align}
\label{metric}
    \text{d}s^2 = -e^{2F_0}Z\text{d}t^2 + e^{2F_1}\left( \frac{\text{d}r^2}{Z} + r^2\text{d}\theta^2 \right)  
	+ e^{2F_2}r^2\sin^2\theta \left(\text{d}\varphi - W\text{d}t\right)^2 \ ,
    \quad
    {\rm with}
    \quad
    Z=1-\frac{r_H}{r}\ ,
\end{align}
where the metric functions $\{F_0,F_1,F_2,W\}$ only depend on the spherical-like coordinates $(r,\theta)$, $t$ is the time coordinate, $\varphi$ is the azimuthal angle, and $r_H>0$ is the radial coordinate of the event horizon $\mathcal{H}$. The latter is a Killing horizon, i.e., there is a Killing vector $\chi$, known as horizon null generator, which is null on $\mathcal{H}$. It reads
\begin{align}
    \chi=\xi+\Omega_H\eta\ ,
    \quad
    \Omega_H=-\left.\frac{g_{t\varphi}}{g_{\varphi\varphi}}\right|_\mathcal{H}=\left.W\right|_\mathcal{H}\ ,
    \label{OmegaH}
\end{align}
where $\Omega_H$ is the horizon angular velocity. 

The complex scalar field is assumed to have the form of Eq.~\eqref{eq:scalar-ansatz}, with $\phi\in\mathbb{R}$. The harmonic time and azimuthal dependencies ensure the energy-momentum tensor is stationary and axisymmetric. The frequency of the scalar field is $\omega>0$ and $m$ is the azimuthal harmonic index, which, as before, can take half-integer values as well~\cite{Pereniguez:2024fkn}. 
%Also, we consider only nodeless solutions ($n=0$).

The ansatz for the Maxwell field is 
\begin{equation}
    A_a\text{d}x^a=(A_t-W A_\varphi)\text{d}t+A_\varphi\text{d}\varphi\ ,
\end{equation}  
in terms of an electric potential $A_t$ and a magnetic potential $A_\varphi$.\footnote{Observe that $A_t$ is not the temporal component of the 4-potential, unless either $W$ or $A_\varphi$ vanish.}

Restricted to this ans\"atze, the EMKG equations result in a set of seven coupled partial differential equations in $r$ and $\theta$. The explicit form of the equations for the metric functions can be found in~\cite{Herdeiro:2015gia}. Together with the appropriate boundary conditions, the system reduces to an elliptic boundary value problem. 

Magnetically neutral hairy BHs within this model reduce to rotating boson stars as the horizon area vanishes~\cite{Delgado:2016jxq}. They are north-south $\mathbb{Z}_2$-symmetric, and the azimuthal harmonic index of the scalar field takes integer values only. Here, following \cite{Pereniguez:2024fkn}, we consider magnetically charged solutions with the same asymptotic leading term as the magnetic KN BH in Eqs.~\eqref{dyonKN1}--\eqref{dyonKN2}, namely with
\begin{equation}
    A_\varphi \to Q_m\cos\theta~~ 
    {\rm as}~~
    r \to \infty\ , 
\end{equation} 
where $Q_m$ is the magnetic charge.

%%%%%%%%%%%%%%%%%%%%%%%%%%%%%%%%%%%%%%%%%%%%%%%%%%%%%%%%%%%%%%%%%%%%%%%%%%%%%%%
\subsection{Boundary conditions and classes of solutions}
\label{bond}
%%%%%%%%%%%%%%%%%%%%%%%%%%%%%%%%%%%%%%%%%%%%%%%%%%%%%%%%%%%%%%%%%%%%%%%%%%%%%%%
To solve the aforementioned system of partial differential we consider the following set of boundary conditions. 
At spatial infinity, the assumption of asymptotic flatness imposes 
\begin{align}
    \left.{F_i}\right|_{r\rightarrow\infty}
    =\left.{W}\right|_{r\rightarrow\infty}
    =0\ . 
\end{align}
On the other hand, the 
asymptotic behaviour of the 
matter fields  reads
%\footnote{In general, $A_t$ can take %any constant value at spatial %infinity.}
% 
\begin{equation}
\label{bcV0}
    \left.{\phi}\right|_{r\rightarrow\infty}=0\ ,
    \quad
    \left.{A_t}\right|_{r\rightarrow\infty}=V_0\ , 
    \quad
    \left.{A_\varphi}\right|_{r\rightarrow\infty}=Q_m\cos \theta~,
\end{equation}
where $V_0$ is a constant. {As for the Maxwell field, this corresponds to a choice of gauge so that $A_t$ approaches a constant value (not necessarily equal to $0$) at spatial infinity. This gauge differs from that used in Section~\ref{section2} when $V_0\neq0$. The gauge used herein can be obtained by applying the gauge transformation function $\alpha=V_0t$ in Eq.~\eqref{eq:gauge} to the scalar and Maxwell fields in Section~\ref{section2}, i.e., $A_t\rightarrow A_t+V_0$ and $\Psi\rightarrow\Psi e^{-ieV_0t}$. Under this gauge transformation, the frequency transforms as $\omega\rightarrow\tilde{\omega}=\omega+eV_0$. In the following, the tilde will be omitted for ease of notation.}

Introducing the radial coordinate $x=\sqrt{r^2-r_H^2}$, $x\in[0,\infty)$, the boundary conditions at the event horizon take a relatively simple form,
\begin{align}
    \left.\partial_x F_i\right|_{x=0}=0\ ,
    \quad
    \left. W\right|_{x=0}=\Omega_H\ ,
    \quad
    \left.\phi\right|_{x=0}=0\ ,
    \quad
    \left.A_t\right|_{x=0}=\Phi_H\ ,
    \quad
    \left. \partial_x A_\varphi \right|_{x=0}=0\ ,
\end{align}
where $\Phi_H$ is the horizon electrostatic potential, respectively. The solutions satisfy the synchronization condition
\begin{align}
\label{synch}
    \omega=m \Omega_H-e\Phi_H\ ,
\end{align} 
where we remark the absence of a magnetic charge contribution. 

In the following, only solutions with $\Phi_H=0$ will be considered to simplify the overall picture. For a trivial scalar field, $\Psi=0$, these are nothing but purely magnetic KN BHs, with vanishing electric charge, $Q_e=0$ (although they possess an induced electric dipole).

Finally, on the symmetry axis, $\theta=0,\pi$, regularity and axisymmetry require 
\begin{align}
    \left.\partial_\theta F_i\right|_{\theta=0,\pi}
    =\left.\partial_\theta W\right|_{\theta=0,\pi}
    =\left.\partial_\theta A_t\right|_{\theta=0,\pi}= 0\ ,
    \quad
    \left.A_\varphi\right|_{\theta=0}=-\left.A_\varphi\right|_{\theta=\pi}=Q_m\ .
\end{align}

Moreover, the absence of conical singularities further requires that $\left.F_1\right|_{\theta=0,\pi}=\left.F_2\right|_{\theta=0,\pi}$. 

As for the scalar field, it is worth noting that the EMKG system (as well as the energy density) includes a term proportional to $ \csc^2\theta (m+eA_\varphi)^2 \phi^2$, which must be finite at $\theta=0,\pi$. A study of an approximate form of the solutions as $\theta \to 0$ and $\theta \to \pi$ reveals the existence of two possible sets of boundary conditions satisfied by the scalar field on the symmetry axis, hence two classes of solutions:
\begin{itemize}
    \item\textbf{Ordinary solutions}. The scalar field satisfies the usual boundary condition
        \begin{align}
            \left.\phi\right|_{\theta=0,\pi}=0\ .
        \end{align}
Such solutions exist for $m\in\mathbb{Z}$ and  $N$ even. They can be regarded as bound states between magnetic monopoles and a BH with gauged scalar hair~\cite{Delgado:2016jxq}. The azimuthal harmonic index $m$ and the number $|N|$ of monopoles are independent parameters in this case.

    \item\textbf{Polar solutions}.
    This case corresponds to
        \begin{align}
           e Q_m=\pm m=\frac{N}{2}\ ,
        \end{align}
        where $m$ can now be a half-integer, $|m|=1/2,1,3/2,\ldots$. 
        The above condition allows for a different set of boundary conditions satisfied by the scalar field amplitude on the symmetry axis such that $ \csc^2\theta (m+eA_\varphi)^2 \phi^2$ is finite there - see Appendix~\ref{App:1}. Restricting (without any loss of generality) to the ``+'' sign, these are
	    \begin{align}
            \label{c11}
            \left.\phi\right|_{\theta=0 }=\left.\partial_\theta \phi\right|_{\theta= \pi }=0\ .
        \end{align}
        That is, the scalar field amplitude has a different behavior at the north and south poles (hence the name ``polar''), and does \textit{not} vanish at $\theta=\pi$. At the linear level, such scalar clouds would describe {\it south monopole modes} \cite{Pereniguez:2024fkn}.
\end{itemize}

%%%%%%%%%%%%%%%%%%%%%%%%%%%%%%%%%%%%%%%%%%%%%%%%%%%%%%%%%%%%%%%%%%
 \subsection{Asymptotics and quantities of interest}
%%%%%%%%%%%%%%%%%%%%%%%%%%%%%%%%%%%%%%%%%%%%%%%%%%%%%%%%%%%%%%%%%%

Most of the physical quantities of interest are encoded in the expression of the metric functions at the event horizon or spatial infinity.
Considering first horizon quantities, we introduce the Hawking temperature $T_H={\kappa}/({2\pi})$, where $\kappa$ is the surface gravity
defined as $\kappa^2=-\frac{1}{2}(\nabla_a \chi_b)(\nabla^a \chi^b)|_\mathcal{H}$, 
and the event horizon area $A_H$. 
These are computed as (see definitions in~\cite{Herdeiro:2015gia})
\begin{eqnarray}
\label{THAH}
&&
T_H=\frac{1}{4\pi r_H}e^{F_0^{(2)}(\theta)-F_1^{(2)}(\theta)} \ ,
\\
&&
A_H=2\pi r_H^2 \int_0^\pi \text{d}\theta\,\sin \theta\,e^{F_1^{(2)}(\theta)+F_2^{(2)}(\theta)} \ .
\end{eqnarray}
The event horizon angular velocity $\Omega_H$ is fixed by the horizon value of the metric function $W$, cf. Eq.~\eqref{OmegaH}.

The ADM mass $M$ and angular momentum $J$ are read from 
the asymptotic sub-leading behavior of the following metric functions:
\begin{eqnarray}
\label{asym}
    g_{tt} =-e^{2F_0}Z+e^{2F_2}W^2r^2 \sin^2 \theta\sim-1+\frac{2GM}{r}+\dots\ ,
    \quad
    g_{t\varphi}=-e^{2F_2}W r^2 \sin^2 \theta\sim-\frac{2GJ}{r}\sin^2\theta+\nonumber \dots\ .  
\end{eqnarray}  
Of interest is also the asymptotic form of the gauge potentials,
\begin{eqnarray}
    \label{asym-matter-fields}
    A_t\sim V_0+\frac{Q_e}{r}+\dots\ ,
    \quad
    A_{\varphi}\sim Q_m \cos\theta+\frac{\mu_m \sin \theta}{r}+\dots,
\end{eqnarray}
where
$Q_e$ and $\mu_m$ result  from numerics,
while $Q_m$ is an input parameter.
Note that the
bound state
condition
for a localized scalar field imposes
\begin{eqnarray}
    \label{condw}
    \mu^2\leq (\omega -e V_0)^2 \ .
\end{eqnarray}
This is different from Eq.~\eqref{bound} because of the gauge choice, as explained in Section~\ref{bond}.

As usual in BH mechanics, the ADM mass and angular momentum can be expressed as a sum of the event horizon and the matter fields contributions, 
\begin{align}
\label{TotalMass}
    M = M_H -2\int_\mathcal{S} \text{d}S_a\bigg({T^a}_b \xi^{b}-\frac{1}{2}T \xi^a \bigg)\ ,
    \quad
    J = J_H + \int_\mathcal{S} \text{d}S_a \left({T^a}_b \eta^{b}-\frac{1}{2}T \eta^a \right) \ ,
\end{align}
where $\mathcal{S}$ is a spacelike surface, bounded by the event horizon $\mathcal{H}$ and the 2-sphere at spatial infinity $\mathcal{S}^2_\infty$.
The horizon contributions are computed as Komar integrals, 
\begin{align}
    M_H = - \frac{1}{8\pi} \oint_\mathcal{H} \text{d}S_{ab} \nabla^a \xi^b\ ,
    \quad
    J_H = \frac{1}{16\pi} \oint_\mathcal{H} \text{d}S_{ab} \nabla^a \eta^b\ .
\end{align}
The horizon mass can also be expressed as
\begin{align}
    M_H &= - \frac{1}{8\pi} \oint_\mathcal{H} \text{d}S_{ab} \left(\nabla^a \chi^b - \Omega_H \nabla^a \eta^b\right)
    =- \frac{1}{8\pi} \oint_\mathcal{H} \text{d}S_{ab} \nabla^a \chi^b+2\Omega_H J_H
    =\frac{\kappa}{4\pi}A_H+2\Omega_H J_H\ .
\end{align}
This is equivalent to
\begin{align}
    M_H = 2 T_H S+2\Omega_H J_H\ ,
\end{align}
where $S=A_H/4$ is the BH entropy. 

%\begin{align}
%    M 
%    &= 2 T_H S + 2\Omega_H J_H - 2\int_\Sigma \text{d}S_a\left({T^a}_b k^{b}-\frac{1}{2}T k^a \right)\\
%    &= 2 T_H S + 2\Omega_H J - 2\Omega_H \int_\Sigma \text{d}S_a \left({T^a}_b m^{b}-\frac{1}{2}T m^a \right) - 2\int_\Sigma \text{d}S_a\left({T^a}_b k^{b}-\frac{1}{2}T k^a \right)
%\end{align}

The ADM mass and angular momentum can be conveniently expressed as
\begin{align}
    M=M_H+M_F+M_\Psi\ ,
    \quad
    J=J_H+J_F+M_\Psi\ ,
\end{align}
where
\begin{align}
    &M_F = -2 \int_\mathcal{S} \text{d}S_a \left[{(T_F)^a}_b \xi^{b}-\frac{1}{2}T_F \xi^a \right]\ ,
    &&J_F = \int_\mathcal{S} \text{d}S_a \left[{(T_F)^a}_b \eta^{b}-\frac{1}{2}T_F \eta^a \right]\ ,\\
    &M_\Psi = -2 \int_\mathcal{S} \text{d}S_a \left[{(T_\Psi)^a}_b \xi^{b}-\frac{1}{2}T_\Psi \xi^a \right]\ ,
    &&J_\Psi = \int_\mathcal{S} \text{d}S_a \left[{(T_\Psi)^a}_b \eta^{b}-\frac{1}{2}T_\Psi \eta^a \right]\ .
\end{align} 

The angular momentum stored in the matter fields can be expressed as the difference of two boundary terms,
\begin{align}
\label{J1}
    J_F+J_{\Psi}
    &=\int_\mathcal{S} \text{d}^3 x \, \sqrt{-g} \, {T^t}_\varphi=\int_\mathcal{S} \text{d}^3 x\,\sqrt{-g} 
\left[F^{ta}F_{\varphi a}+(D^t\Psi)^*(D_\varphi\Psi)+(D_\varphi\Psi)^*(D^t\Psi)\right]\nonumber\\
    &=\int_\mathcal{S} \text{d}^3 x\,\sqrt{-g}\,\nabla_a\left[\left(\frac{m}{e}+A_\varphi\right)F^{at}\right]
    =\oint_{\mathcal{S}^2_\infty} \text{d}S_r\left(\frac{m}{e}+ A_\varphi\right) F^{rt}
-\oint_\mathcal{H} \text{d}S_r\left(\frac{m}{e}+ A_\varphi\right) F^{rt}\ ,
\end{align}  
with
\begin{eqnarray}
\oint_{\mathcal{S}^2_\infty} \text{d}S_r \left(\frac{m}{e}+ A_\varphi\right) F^{rt}=\frac{4 \pi mQ_e}{e} \ .
\end{eqnarray} 

%%%%%%%%%%%%%%%%%%%%%%%%%%%%%%%%%%%%%%%%%%%%%%%%%%%%%%%%%%%%%%%%%%%%%%%%%%%%%%%
\subsection{Scalings and input parameters} 
%%%%%%%%%%%%%%%%%%%%%%%%%%%%%%%%%%%%%%%%%%%%%%%%%%%%%%%%%%%%%%%%%%%%%%%%%%%%%%%
%
In the numerics, we take 
\begin{equation}
    \Psi \to \Psi/{\sqrt{G}}\ ,
    \quad
    A \to A/{\sqrt{G}}\ ,
    \quad
    r \to r\mu\ ,
    \quad
    \omega \to \omega/\mu\ ,
    \quad
    e\to e {\sqrt{G}} /\mu\ ,
    \quad
    Q_m \to Q_m \mu\ ,
\end{equation}
i.e., we work in units with $G=\mu=1$.

We are  mainly interested in polar solutions satisfying the condition
\begin{equation}
    \label{condition}
    m=q\ ,
\end{equation}
which  allow for a \textit{nonvanishing} scalar field at $\theta=\pi$ (while $\phi=0$ at $\theta=0$). This is a new feature, absent in other BHs with scalar hair discussed in the literature.

Naively, one may anticipate that polar solutions result in stronger violations of the $\mathbb{Z}_2$-symmetry than ordinary solutions. This expectation is contradicted by numerical results.
From our analysis, polar solutions exist only in a small domain close to the existence line, implying they do not carry a significant amount of hair, and the violation of the $\mathbb{Z}_2$-symmetry is rather small. Larger deviations, as those observed in the solutions studied in the next section, were found for ordinary solutions, with a sufficiently large azimuthal harmonic index $m$.

%%%%%%%%%%%%%%%%%%%%%%%%%%%%%%%%%%%%%%%%%%%%%%%%%%%%%%%%%%%%%%%%%%%%%%%%%%%%%%%
\subsection{Results}
%%%%%%%%%%%%%%%%%%%%%%%%%%%%%%%%%%%%%%%%%%%%%%%%%%%%%%%%%%%%%%%%%%%%%%%%%%%%%%%

We have investigated more thoroughly \textit{polar solutions}, which are qualitatively new as compared to other synchronized BHs, being supported by the presence of magnetic monopoles.
Some \textit{ordinary solutions} have also been constructed. However, the study in that case is far less systematic.  

All solutions we report satisfy the synchronization condition in Eq.~~\eqref{sync} with $\Phi_H=0$, since we have imposed $\left.A_t\right|_{r=r_H}=0$. As for the electric potential boundary conditions at spatial infinity, we have considered two cases.

First, we have the \textbf{zero electric potential solutions} ($V_0=0$), which are found
by imposing 
\begin{eqnarray}
\label{c1}
    \lim_{r\to \infty} A_t=0\ ,
\end{eqnarray}
in which case the (electric) chemical potential vanishes. This choice of boundary condition is mainly motivated by the fact that the purely magnetic KN BHs
also satisfy it. However, their electric charge vanishes, which is not the case for the hairy solutions.

Second, the \textbf{zero electric charge solutions} ($Q_e=0$) are found by imposing the boundary condition
\begin{eqnarray}
\label{c2}
    \lim_{r\to \infty}r^2 \partial_r A_t=0\ ,
\end{eqnarray}
in which case
the value $V_0$ of the electric potential at infinity is non-zero and is read from the numerical output. This choice is mainly motivated by the fact that
the superradiance endpoints of purely magnetic KN BHs are likely to be among the latter configurations, assuming that the growth and saturation of superradiant instabilities are nearly conservative.

Both solutions, with either $V_0=0$ or $Q_e=0$, share the same existence line. The following remarks apply to both cases, always for polar solutions, unless otherwise mentioned.

Fixing $\{Q_m, m, \omega \}$, families of solutions are found by varying the event horizon radius, $r_H$. In all cases studied herein, the solutions form a sequence starting on the corresponding existence line, in which the scalar field vanishes and a particular purely magnetic KN BH is approached. These hairy BHs can be regarded as non-linear continuations of the scalar clouds discussed in Section~\ref{section2}. 

\begin{figure}[h!]
\centering
    \includegraphics[width=0.45\textwidth]{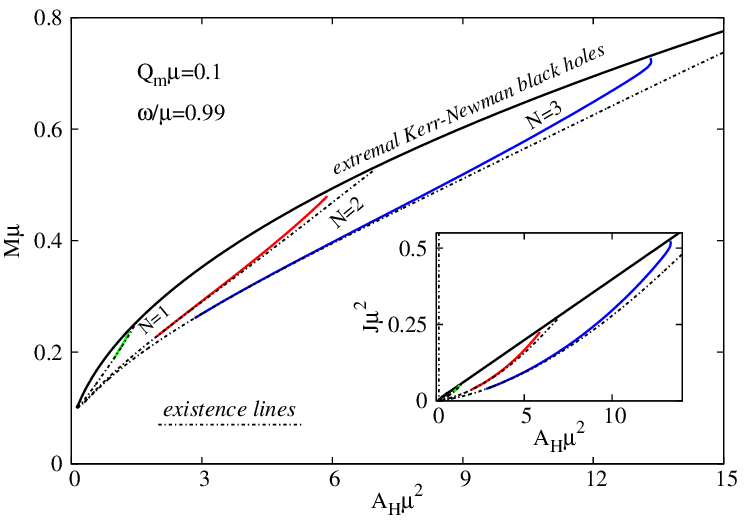} 
    \includegraphics[width=0.45\textwidth]{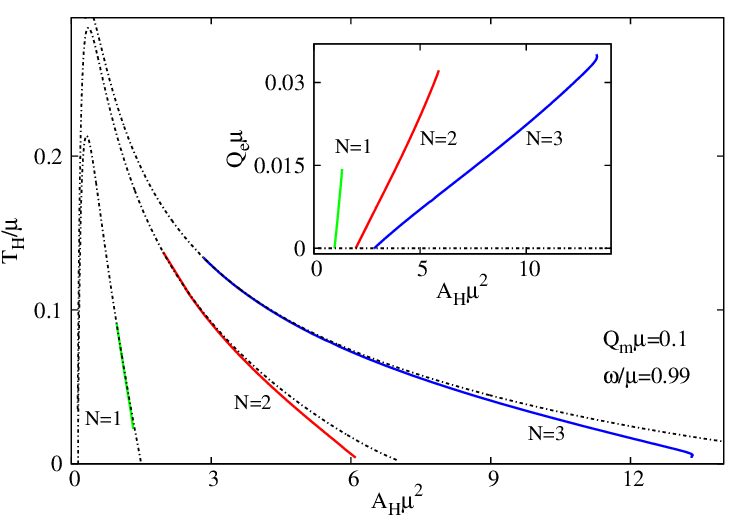} 
    \caption{Sequences of polar solutions, with $V_0=0$ and fixed $\{Q_m\mu,\omega/\mu\}$, for magnetic hairy BHs with $N=1,2,3$. The solutions bifurcate from the corresponding points on the existence line and end 
in extremal hairy BHs.} 
\label{data}
\end{figure}

Sequences of solutions with  $V_0=0$ and different numbers of monopoles $|N|$ are shown in Fig.~\ref{data} in an ADM mass (left panel) or temperature (right panel) vs. horizon area plot. 
One can notice that they end in critical configurations which possess zero Hawking temperature and finite horizon area and global charges. The whole domain of existence can in principle be scanned by varying 
$\{r_H,\omega\}$. Although the scanning of the parameter space is so far rather limited, the picture above is likely to be universal. In particular, hairy BHs exist for a limited range of frequencies $0<\omega_\text{min}<\omega<\omega_\text{max}=\mu$, and, as such, they do not possess a static limit, as for other families of BHs with synchronized scalar hair. However, as opposed to them, the solutions herein do not possess a solitonic limit. Besides, just like similar hairy BHs, violations of the KN-bound $J^2/M^4+(Q_e^2+Q_m^2)/M^2 \leq 1 $ are found - see, e.g.,~\cite{Delgado:2016zxv}.

As mentioned above,
a curious feature of the solutions 
with 
$V_0=0$
is that they possess
a non-zero electric charge $Q_e$, despite the fact that 
the electrostatic potential vanishes ($\Phi_H=0$) - see Fig.~\ref{data} (right panel, inset).

 %%%%%%%%%%%%%%%%%%%%%%%%%%%%%%%%%%%%%%%%% 

  %%%%%%%%%%%%%%%%%%%%%%%%%%%%%%%%%%%%%%%%% 
\begin{figure}[h!]
\begin{center}
\includegraphics[width=0.45\textwidth]{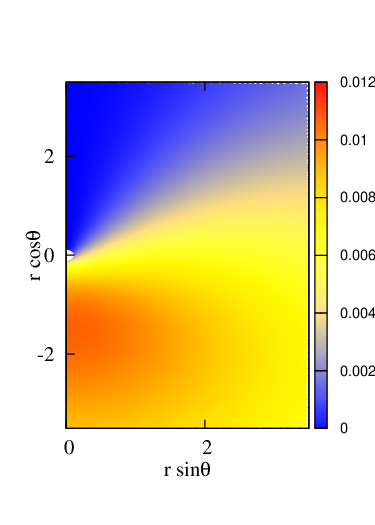} 
\includegraphics[width=0.45\textwidth]{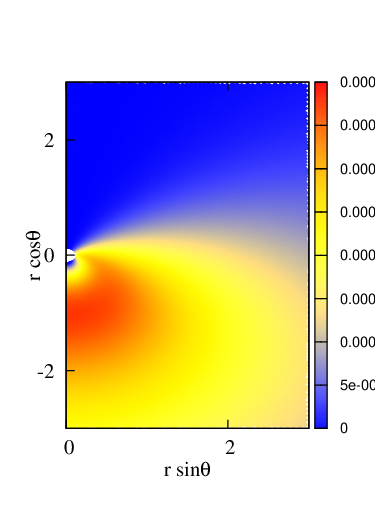} 
\caption{  
The  scalar field (left) and the Komar mass density
(right)
are shown for a solution
with $N=2$, $Q_m\mu=0.25$, $r_H\mu=0.125$ and $\omega/\mu=0.994$. 
}
\label{profile}
\end{center}
\end{figure} 
%%%%%%%%%%%%%%%%%%%%%%%%%%%%%%%%%%%%%%%%%%%

The profile of the scalar field amplitude together with the scalar field contribution to the Komar mass density are shown as color maps in Fig. \ref{profile}. 
%\ch{[No such figure.]}
The violation of the $\mathbb{Z}_2$-symmetry and the absence of an equatorial plane are obvious.
Naively, this should lead to a significant violation of the $\mathbb{Z}_2$-symmetry on the geometry side. However, this is not the case, at least for the solutions obtained so far. This can be explained by noticing that the maximal contribution of the scalar field to the ADM mass is at most several percent, achieved by the extremal configurations. That is, for all studied \textit{polar solutions}, most of the mass and angular momentum outside the horizon is stored in the electromagnetic field. The situation is different for \textit{ordinary solutions}, in which case hairier solutions have been obtained (even without a systematic investigation of the parameter space).

An interesting feature of the solutions with $V_0=0$
is that the scalar field does not trivialize as 
$\omega \to \mu$.
Instead, a set of configurations describing 
\textit{marginal bound states}
arises in that limit.
 The scalar field is still localized,
 and the global charges are still finite.
The picture is different for solutions with $Q_e=0$. The limit $\omega\to\mu$ is not approached for the cases considered herein.

\begin{figure}[h!]
\centering
    \includegraphics[width=0.45\textwidth]{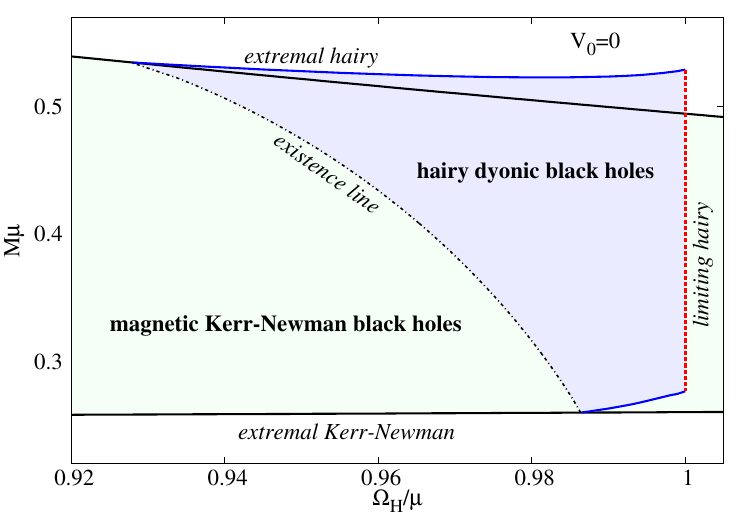} 
    \includegraphics[width=0.45\textwidth]{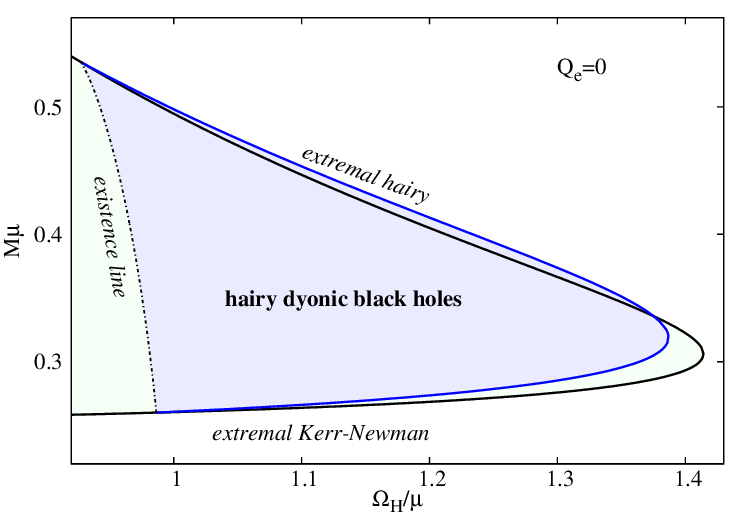} 
    \caption{Domain of existence  of 
    polar solutions  with  
    $N=2$, $Q_m\mu=0.25$
   is shown for 
    a vanishing 
    electrostatic potential
    (left panel)
    and a vanishing electric charge
    (right panel). 
    }

    \label{domain}
\end{figure}

Based on the existent numerical results,
one can conjecture that,
given a number $|N|$ 
of magnetic monopoles
together with a value of the gauge coupling constant $e$,
the domain of existence of hairy BHs
form a single, compact region - 
see Fig.~\ref{domain}.
In the case $V_0=0$,
this region is bounded by three curves: ($i$) the existence line, ($ii$) the set of marginally bound states (satisfying $\omega=\mu$), and ($iii$) the set of zero-temperature, extremal BHs.
This last set of solutions consists of two lines: each extremal line bifurcates from the corresponding Hod point and ends in a marginally bound solution with $T_H=0$. 
The picture is different
for the solutions
with $Q_e=0$, 
the marginally bound states being absent in this case,
while the curve of extremal solutions connects two Hod points.

 %%%%%%%%%%%%%%%%%%%%%%%%%%

 %%%%%%%%%%%%%%%%%%%%%%%%%%
\begin{figure}[h!]
\centering
    \includegraphics[width=0.45\textwidth]{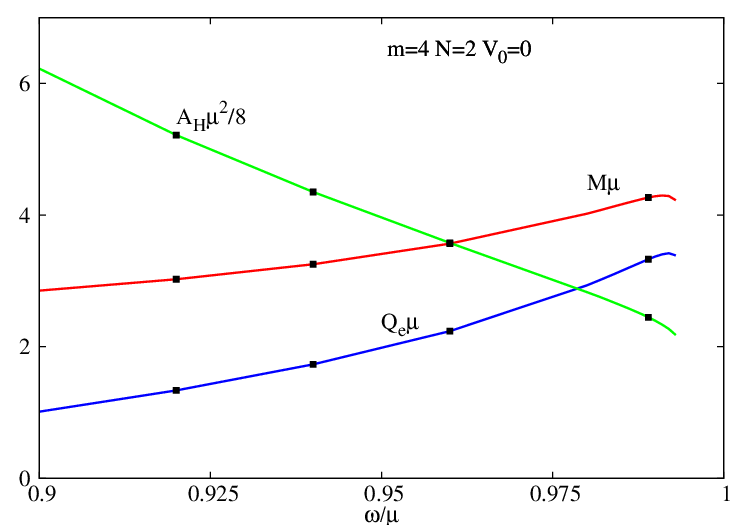} 
    \includegraphics[width=0.45\textwidth]{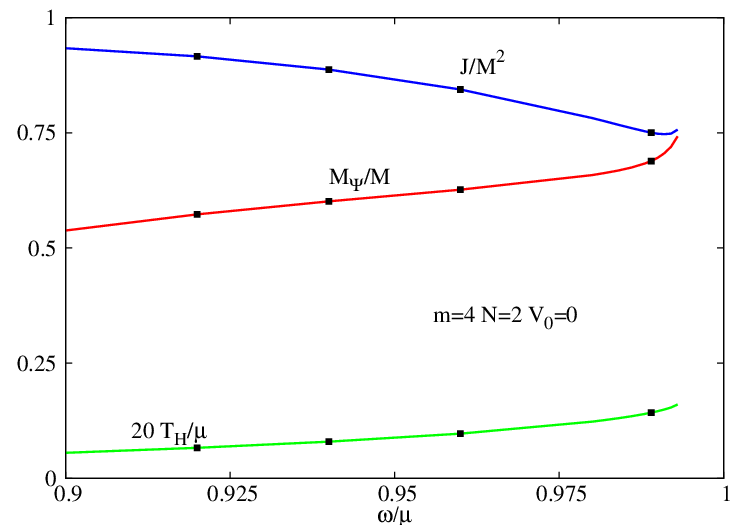} 
    \caption{Some physical quantities of a sample of \textit{ordinary solutions} with fixed $r_H\mu=0.14$, $Q_m\mu=0.9$, $V_0=0$, $N=2$ and $m=4$.
    The points indicate the four configurations considered in Fig.~\ref{fig:shadows}.
    }

    \label{ordinary}
\end{figure} 

 %%%%%%%%%%%%%%%%%%%%%%%%%%

Our study of \textit{ordinary solutions} has been much less systematic. Still, the analysis done so far indicates that such solutions with $V_0=0$
  share their basic properties with the \textit{polar} ones.
   For example, they also exist for $0<\omega_\text{min}<\omega<\omega_\text{max}=\mu$,    with the absence of a
   static  limit.  Some results for a family of
  {\it ordinary solutions} 
  with  fixed values of the input
  parameters
  $\{m,N,Q_m,r_H\}$
  and
  $V_0=0$
  are shown in Fig.~\ref{ordinary}. These include the solutions for which the shadows and gravitational lensing will be analyzed in the next section.

%%%%%%%%%%%%%%%%%%%%%%%%%%%%%%%%%%%%%%%%%%%%%%%%%%%
% SECTION
%%%%%%%%%%%%%%%%%%%%%%%%%%%%%%%%%%%%%%%%%%%%%%%%%%%
%%%%%%%%%%%%%%%%%%%%%%%%%%%%%%%%%%%%%%%%%%%%%%%%%%%
\section{Shadows and gravitational lensing}
\label{section5}
%%%%%%%%%%%%%%%%%%%%%%%%%%%%%%%%%%%%%%%%%%%%%%%%%%%
%%%%%%%%%%%%%%%%%%%%%%%%%%%%%%%%%%%%%%%%%%%%%%%%%%%

The direct observation by the Event Horizon Telescope (EHT) of the supermassive BH candidates M87* and SgrA* has transformed the study of strong gravitational lensing into an active research field. The study of lensing properties of compact objects serves as a diagnosis of their strong gravity imprints - see, e.g.,~\cite{Cunha:2018acu} for a review. 
In this section, we very briefly explore a few lensing images of magnetically charged BHs with synchronized gauged scalar hair and discuss how some lensing features might be shared generically with other solutions. As previously mentioned, polar solutions often do not carry a significant amount of hair and the violation of the $\mathbb{Z}_2$-symmetry is rather small. For this reason, in this section we shall focus on ordinary solutions, in particular those highlighted in Fig.~\ref{ordinary}, which can display much larger deviations from KN BHs, given a large enough value of the azimuthal harmonic index $m$.

The observation image of a compact object, for a given observer, can be simulated numerically via {\it backward ray-tracing}. Under this procedure, each pixel of the synthetic image corresponds to a slightly different observation direction in the observer's local sky, and contains the information of the light ray received along that direction. The pixel information can be modeled by propagating null geodesics backward in time, starting from the observation location along the detection direction, until a light source can be found along the geodesic path - see, e.g.,~\cite{Bohn:2014xxa,Cunha:2016bpi} for details. 
We shall consider a very large (far-away) colored sphere as the light source at a constant radial coordinate, such that the sphere contains both the BH and the observer inside it. The sphere is divided into four colored quadrants (red, yellow, green, blue), superimposed with a black mesh, following~\cite{Bohn:2014xxa,Cunha:2015yba, Cunha:2018gql,Cunha:2018cof}. We shall further assume an optically transparent medium between the observer and the light sources. This academic setup neatly illustrates how different image pixels are mapped into the far-away sphere.

\begin{figure}[h!]
\centering
    \includegraphics[width=0.35\textwidth]{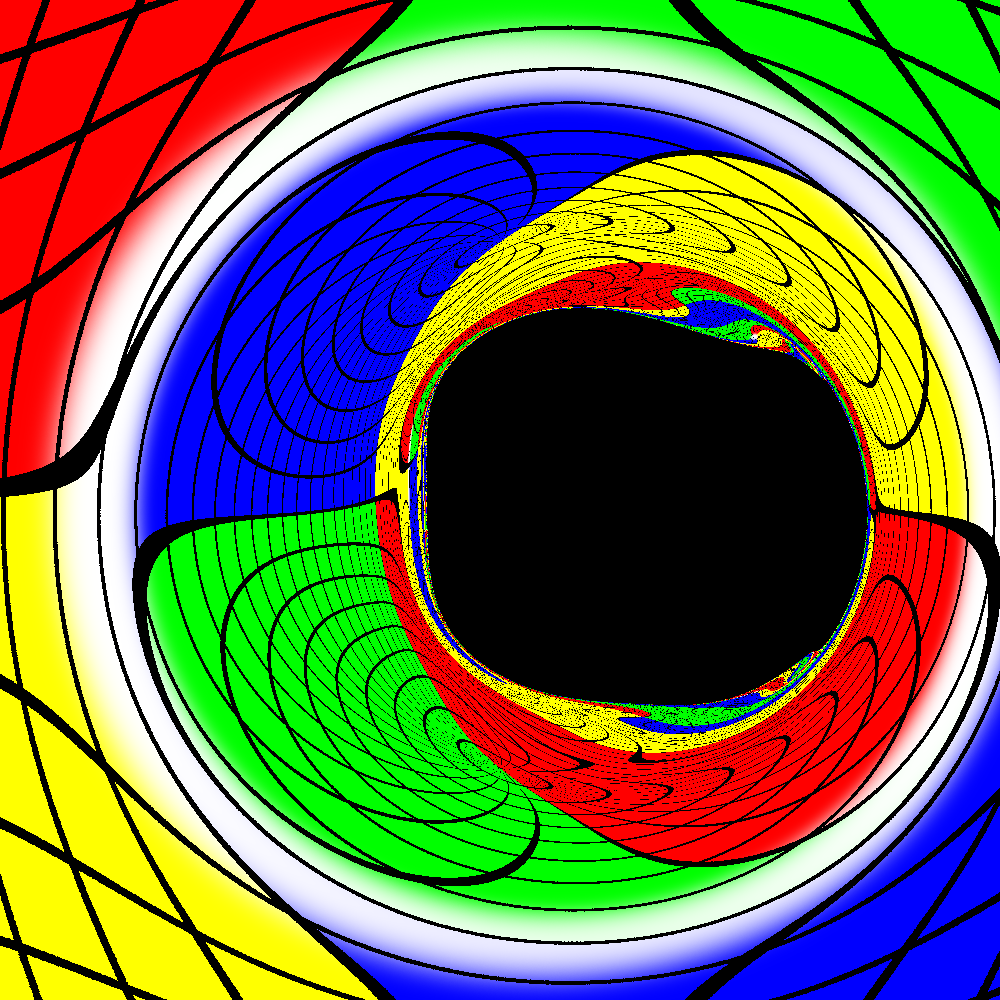}\includegraphics[width=0.35\textwidth]{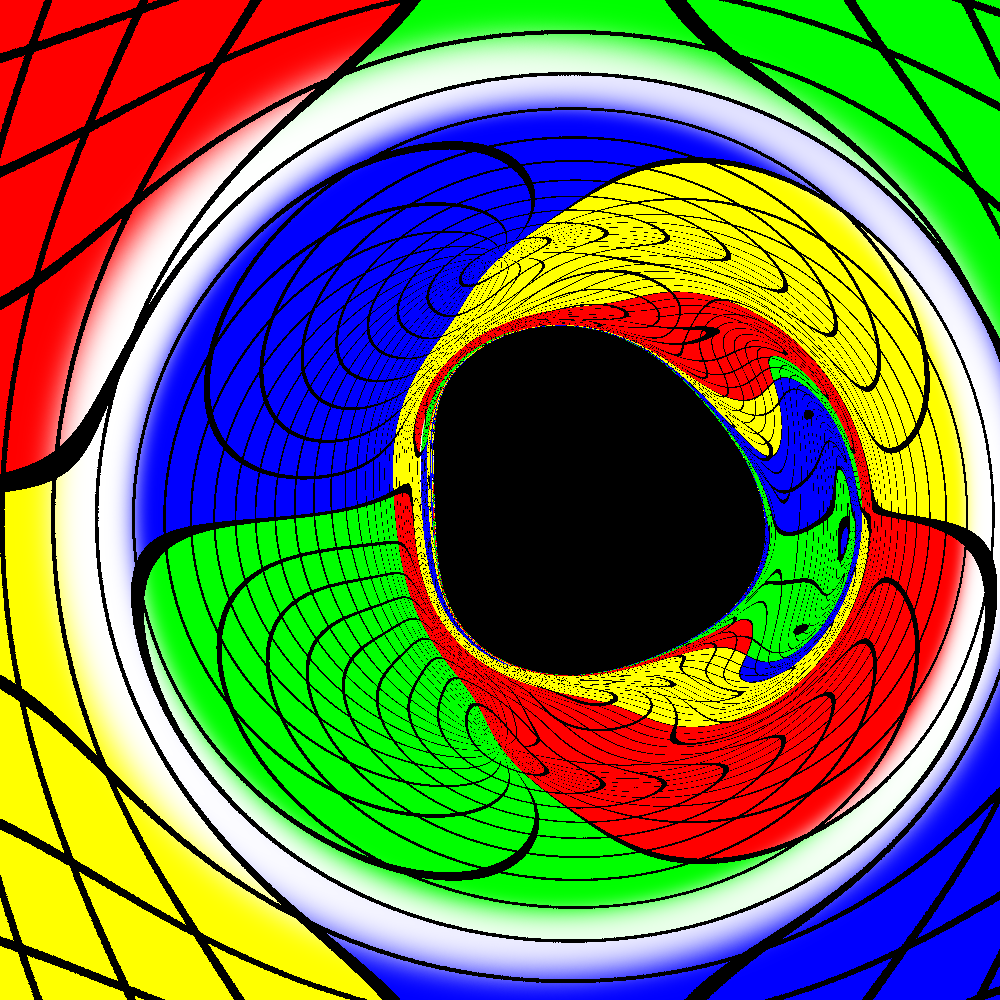} 
    \includegraphics[width=0.35\textwidth]{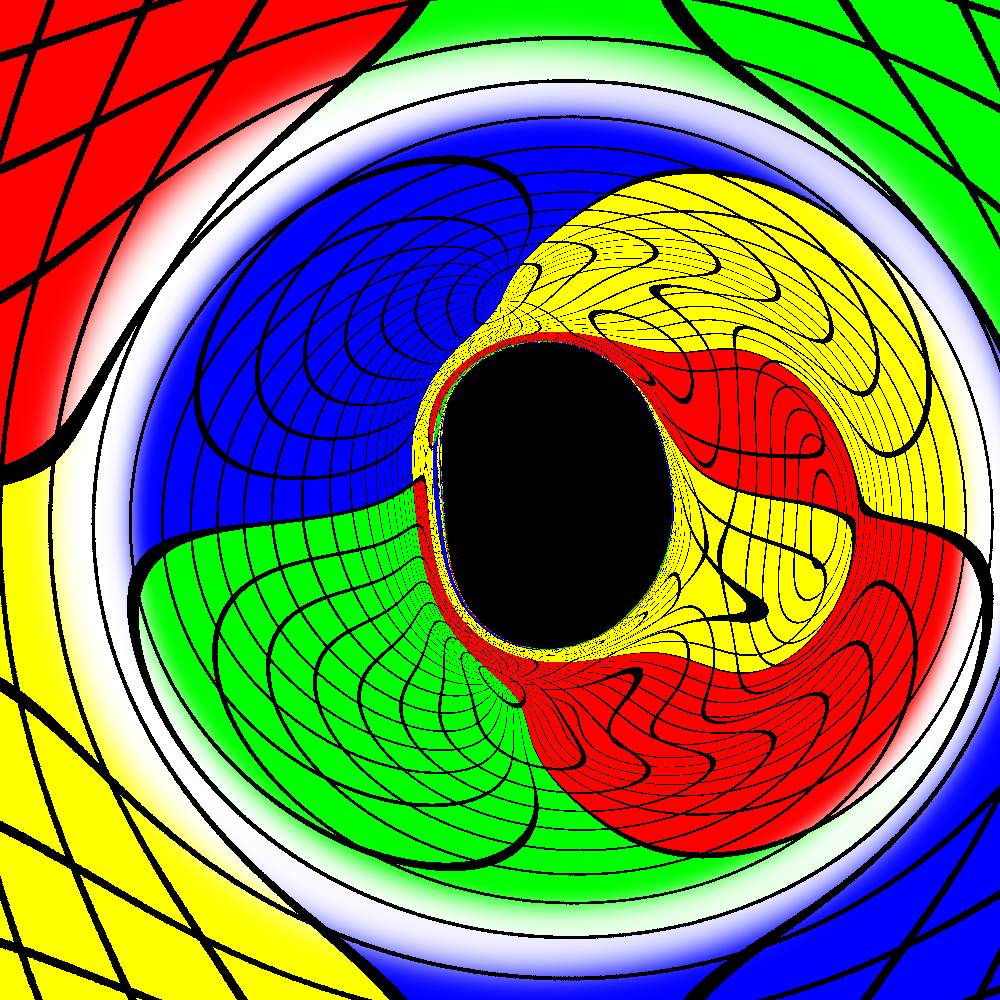}\includegraphics[width=0.35\textwidth]{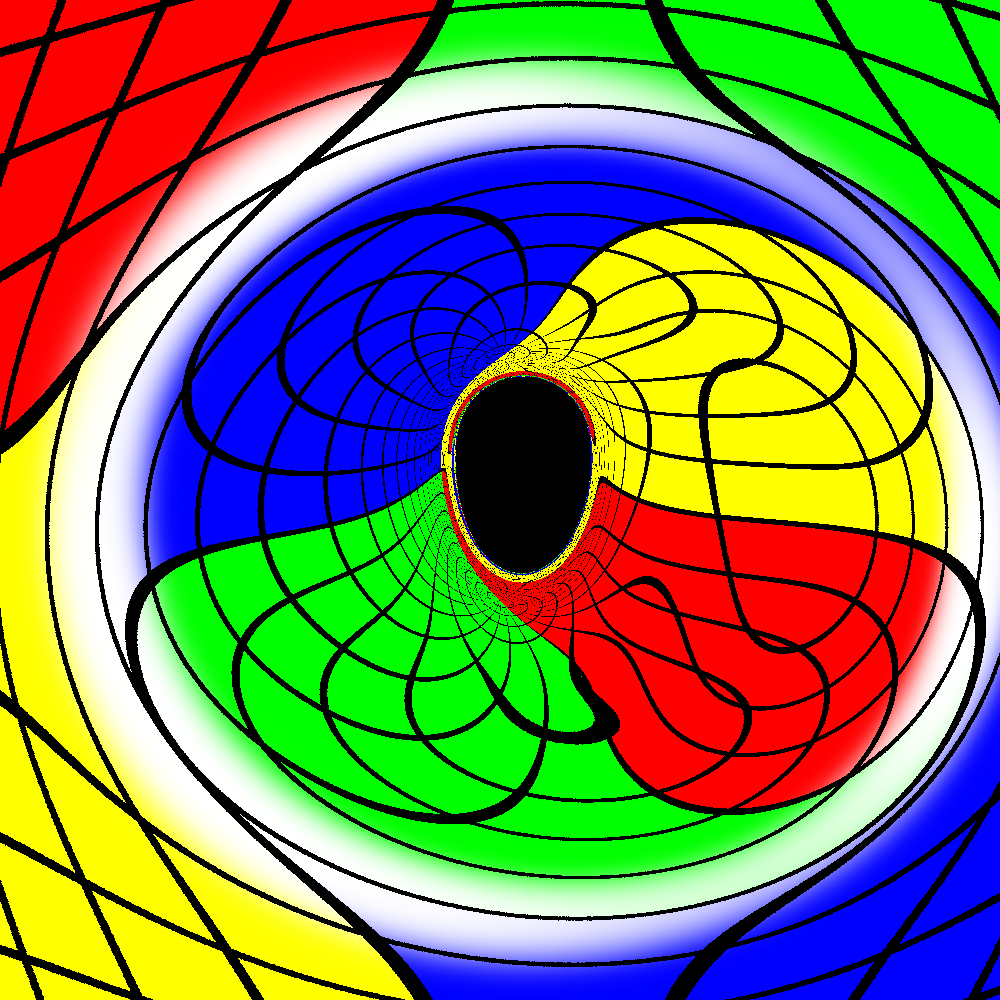}
    \caption{Shadows images and gravitational lensing of ordinary magnetic hairy BHs with $r_H\mu=0.14$, $V_0=0$, and with harmonic frequency (from left to right and from top to bottom) $\omega/\mu=\{0.92,0.94,0.96,0.989\}$.} 
\label{fig:shadows}
\end{figure}

The observation images of a few ordinary hairy BH solutions are displayed in Fig.~\ref{fig:shadows}. A noticeable feature in all the images is the signature of a violation of the north-south $\mathbb{Z}_2$-symmetry. This is clear because the shadows are not invariant under reflection along the horizontal axis of the image.  This asymmetry is present even though the observer was placed in the equatorial plane, at $\theta = \pi/2$. As $\omega/\mu$ increases, the shadow (the dark region in the image) becomes more noticeably distorted as compared to the KN case. The bottom-right image in particular displays strong gravitational lensing with a fairly small shadow. The latter suggests a more significant amount of energy is being stored in the gauged scalar field, which can be confirmed in Fig.~\ref{ordinary}.

%%%%%%%%%%%%%%%%%%%%%%%%%%%%%%%%%%%%%%%%%%%%%%%%%%%
% SECTION
%%%%%%%%%%%%%%%%%%%%%%%%%%%%%%%%%%%%%%%%%%%%%%%%%%%
%%%%%%%%%%%%%%%%%%%%%%%%%%%%%%%%%%%%%%%%%%%%%%%%%%%
\section{Conclusion}
\label{section6}
%%%%%%%%%%%%%%%%%%%%%%%%%%%%%%%%%%%%%%%%%%%%%%%%%%%
%%%%%%%%%%%%%%%%%%%%%%%%%%%%%%%%%%%%%%%%%%%%%%%%%%%
In this paper, we have considered how to add hair to a KN BH with a magnetic (and possibly electric) charge. The existence of a magnetic charge allows a novel feature for the stationary scalar clouds that can be in equilibrium with the BH, under the synchronization condition, in a test field approximation: they break the north-south $\mathbb{Z}_2$-symmetry of the background KN BH solution. Then, upon backreacting these clouds, a family of hairy BHs for which the geometry breaks the north-south $\mathbb{Z}_2$-symmetry is found.

The north-south asymmetric hairy BHs constructed herein will exhibit different phenomenological properties. Here, we only scratched the surface by briefly considering the shadows and gravitational lensing of some ordinary solutions. Studies of timelike geodesics, for instance, could certainly be of interest.

Finally, a more challenging goal would be to study the dynamical formation of such BHs from superradiance. In this respect, the breakdown of the north-south symmetry may lead to non-trivial features, e.g., a kick for the final BH due to the shift of the center of mass. Analyzing this possibility in full numerical relativity may be within reach of current infrastructures.

\begin{acknowledgments}
This work is supported by the Center for Research and Development in Mathematics and Applications (CIDMA) through the Portuguese Foundation for Science and Technology (FCT -- Fundaç\~ao para a Ci\^encia e a Tecnologia) through projects: UIDB/04106/2020 (with DOI identifier \url{https://doi.org/10.54499/UIDB/04106/2020}); UIDP/04106/2020 (DOI identifier \url{https://doi.org/10.54499/UIDP/04106/2020});  PTDC/FIS-AST/3041/2020 (DOI identifier \url{http://doi.org/10.54499/PTDC/FIS-AST/3041/2020});  and 2022.04560.PTDC (DOI identifier \url{https://doi.org/10.54499/2022.04560.PTDC}). This work has further been supported by the European Horizon Europe staff exchange (SE) programme HORIZON-MSCA-2021-SE-01 Grant No.\ NewFunFiCO-101086251.
P.C. is supported by the Individual CEEC program \url{http://doi.org/10.54499/2020.01411.CEECIND/CP1589/CT0035} of 2020, funded by FCT. 
 Computations have been performed at the Argus
cluster at the U.~Aveiro.
\end{acknowledgments}

\appendix

\section{Monopole spherical harmonics}
\label{App:1}
 
% Dyonic Kerr--Newman BHs are subject to superradiant energy extraction by bosonic fields~\cite{Pereniguez:2024fkn}. In fact, there can be both rotation driven superradiance, which occurs for the uncharged Kerr background, and charge driven superradiance, which occurs for the unspinning, Reissner-Nordstr\"om (dyonic) background. Considering the latter case, one can appreciate a qualitatively new feature due to the presence of the magnetic charge $Q_m$.

%On the other hand, when the scalar field is gauged ($e\neq0$), $Y(\theta,\varphi)=Y_{q,\ell,m}(\theta,\varphi)=\Theta_{q,\ell,m}(\theta)e^{im\varphi}$, where $Y_{q,\ell,m}$ denotes monopole spherical harmonics~\cite{Wu:1976ge}, with $q\equiv eQ_m$, as defined in the Appendix~\ref{App:1}. 

\begin{figure}[h]
    \centering
    \begin{minipage}[b]{.12\linewidth}
        \centering
        \includegraphics[width=\columnwidth]{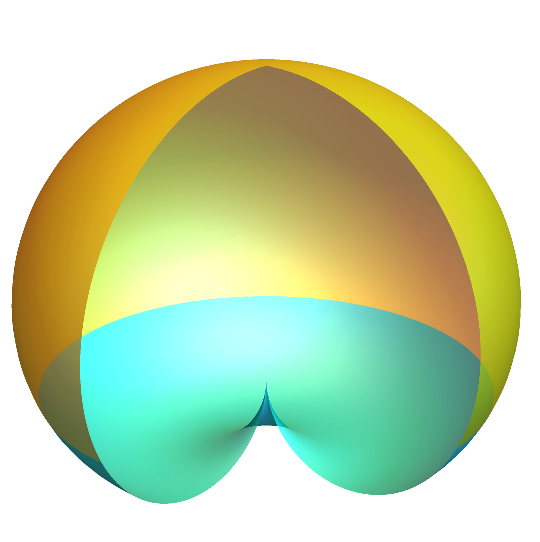}

        \medskip
        
        $\ell=\dfrac{1}{2}$
    \end{minipage}
    \qquad
    \begin{minipage}[b]{.12\linewidth}
        \centering
        \includegraphics[width=\columnwidth]{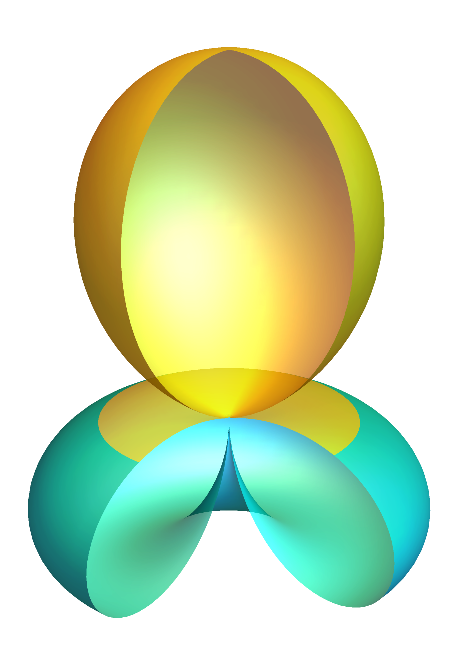}
        
        \medskip
        
        $\ell=\dfrac{3}{2}$
    \end{minipage}
    \qquad
    \begin{minipage}[b]{.12\linewidth}
        \centering
        \includegraphics[width=\columnwidth]{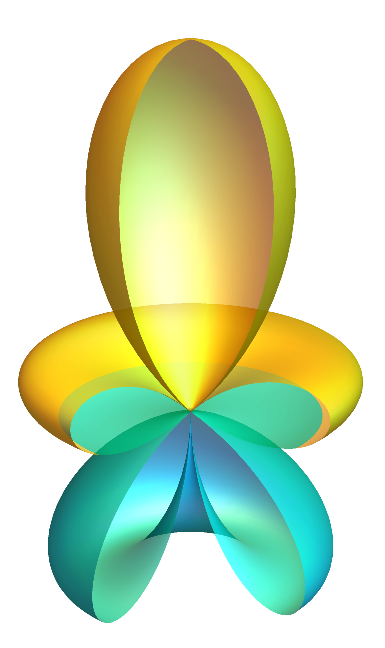}
        
        \medskip
        
        $\ell=\dfrac{5}{2}$
    \end{minipage}
    \qquad
    \begin{minipage}[b]{.12\linewidth}
        \centering
        \includegraphics[width=\columnwidth]{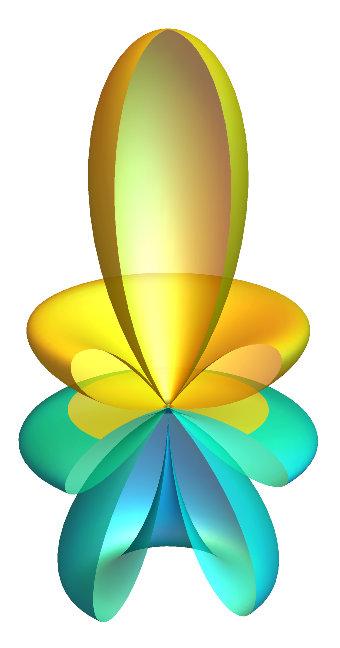}
        
        \medskip
        
        $\ell=\dfrac{7}{2}$
    \end{minipage}
    \qquad
    \begin{minipage}[b]{.12\linewidth}
        \centering
        \includegraphics[width=\columnwidth]{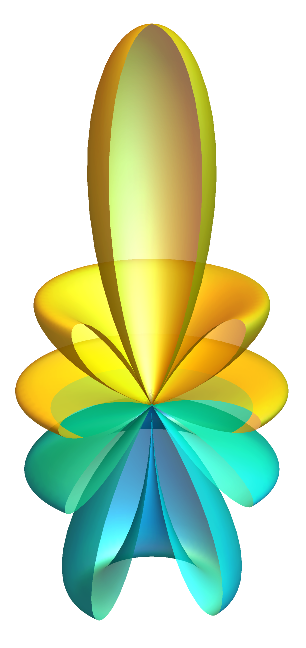}
        
        \medskip
        
        $\ell=\dfrac{9}{2}$
    \end{minipage}
    \qquad
    \begin{minipage}[b]{.12\linewidth}
        \centering
        \includegraphics[width=\columnwidth]{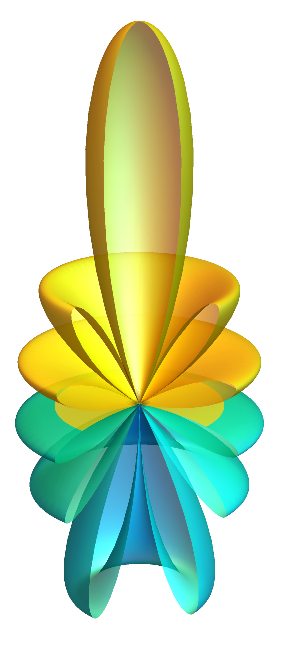}
        
        \medskip
        
        $\ell=\dfrac{11}{2}$
    \end{minipage}
    \caption{Low-$\ell$ monopole spherical harmonics with $-q=m=1/2$. Three-dimensional plots of spherical radius $|Y_{q,\ell,m}(\theta,\varphi)|$, where $\theta\in[0,\pi]$ and $\varphi\in[0,2\pi)$ are the usual spherical coordinates. For ease of visualization, $\varphi$ varies from $0$ to $3\pi/2$. Additionally, the north (south) hemisphere is shaded yellow (blue).}
    \label{fig:10}
\end{figure}

%
The monopole spherical harmonics $Y_{q,\ell,m}$ are defined as~\cite{Wu:1976ge}
\begin{align}
    Y_{q,\ell,m}(\theta,\varphi)=\Theta_{q,\ell,m}(\theta)e^{im\varphi}=N_{q,\ell,m}(1-u)^\frac{|\alpha|}{2}(1+u)^\frac{|\beta|}{2}P_\nu^{(|\alpha|,|\beta|)}(u)e^{im\varphi}\ ,
\end{align}
where $\ell=|q|,|q|+1,\ldots$, $m=-\ell,-\ell+1,\ldots,\ell-1,\ell$, $u=\cos\theta$, $P_\nu^{(|\alpha|,|\beta|)}$ denote the Jacobi polynomials, with
\begin{align}
    &\alpha=-q-m\ ,
    \quad
    \beta=q-m\ ,
    \quad
    \nu=\ell+m+\frac{\alpha-|\alpha|+\beta-|\beta|}{2}\ ,\\
    &N_{q,\ell,m} =\frac{(-1)^{\frac{\alpha-|\alpha|}{2}}}{\sqrt{4\pi}}\sqrt{\frac{2\ell+1}{2^{|\alpha|+|\beta|}}\frac{\nu!(\nu+|\alpha|+|\beta|)!}{(\nu+|\alpha|)!(\nu+|\beta|)!}}\ .
\end{align}
The first few monopole spherical harmonics with $-q=m=1/2$ are illustrated in Fig.~\ref{fig:10}.

They satisfy the orthonormality relation
\begin{align}
    \int_0^{2\pi}\text{d}\varphi\int_0^\pi \text{d}\theta\,\sin\theta\,Y^*_{q,\ell',m'}(\theta,\varphi)Y_{q,\ell,m}(\theta,\varphi) = \delta_{\ell'\ell}\delta_{m'm}\ ,
\end{align}
where $\delta_{mn}$ is the Kronecker delta. 

The monopole spherical harmonics reduce to ordinary spherical harmonics $Y_{\ell,m}$ when $q=0$. At the north (south) pole, where $\theta=0$ ($\theta=\pi$) and $\varphi$ is undefined, monopole spherical harmonics vanish unless $q\pm m=0$. In other words, when $q\pm m=0$, $Y_{q,\ell,m}$ is not well-defined at $\cos\theta=\pm 1$, which implies that the radial function in Eq.~\eqref{eq:scalar-ansatz} must vanish and we only have trivial mode solutions. However, we can use the gauge invariance in Eq.~\eqref{eq:gauge} to get non-trivial solutions. In fact, taking the gauge transformation function $\alpha=\alpha_\pm\equiv\pm Q_m\varphi$, the angular dependence of the scalar field transforms as
\begin{align}
    Y_{q,\ell,m}(\theta,\varphi)\rightarrow Y_{q,\ell,m}^\pm(\theta,\varphi)=\Theta_{q,\ell,m}(\theta)e^{i(q\pm m)\varphi}\ .
\end{align}
Then, if $q\pm m=0$, $Y_{q,\ell,m}^\pm$ does not need to vanish so that the solution is well-defined at the poles. In fact, when $q\pm m=0$, monopole spherical harmonics are non-zero and finite at $\cos\theta=\pm1$ and vanish at $\cos\theta=\mp 1$. As opposed to the ungauged case, the values of $m$ are not restricted to integers, but can also be half-integers (as long as $q\pm m\in\mathbb{Z}$).

Additionally, note that spherical harmonics are $\mathbb{Z}_2$-symmetric, up to sign, with respect to the equatorial plane, $\theta=\pi/2$. In particular, $Y_{\ell,m}(\pi-\theta,\varphi)=(-1)^{\ell+m}Y_{\ell,m}(\theta,\varphi)$ and, thus, $Y_{\ell,m}$ is an even (odd) function with respect to the plane $\theta=\pi/2$ whenever $\ell+m$ is even (odd). As for monopole spherical harmonics, since $Y_{q\,\ell,m}(\pi-\theta,\varphi)=(-1)^{\ell-q-|q+m|} Y_{-q,\ell,m}(\theta,\varphi)$, the $\mathbb{Z}_2$ symmetry can be broken in this case. 

\section{Stationary scalar clouds: extremal case}
\label{App:2}
Consider extremal dyonic KN BHs, for which $M^2=a^2+Q_e^2+Q_m^2$ and $\Delta=(r-M)^2$. Defining
\begin{align}
    R(r)=\frac{MW(r)}{r-M}\ ,
\end{align}
and introducing a new (dimensionless) independent variable
\begin{align}
    x\equiv\frac{r-M}{M}\ ,
\end{align}
the radial equation becomes
\begin{align}
    &\left[-\frac{\text{d}^2}{\text{d}x^2}+\frac{2M^2\mu^2-2eQ_eM\omega_c-4M^2\omega_c^2}{x}+\frac{p^2-\frac{1}{4}}{x^2}\right]W(x)=(\omega_c^2-\mu^2)M^2W(x)\ ,
\end{align}
where
\begin{align}
    p^2=\Lambda+\mu^2M^2-e^2Q_e^2-q^2-a^2\omega_c^2-6M^2\omega_c^2-6eQ_eM\omega_c+\frac{1}{4}\ .
\end{align}
This resembles the radial part of an hydrogen-like Schr\"{o}dinger equation, whose potential depends on the ``energy'' $\omega_c^2$. Note that $\omega_c^2-\mu^2$ is negative if $\omega_c^2<\mu^2$ (bound states) and positive if $\omega_c^2>\mu^2$ (scattering states). We are interested in the former, here dubbed stationary clouds. Considering
\begin{align}
    \epsilon\equiv M\sqrt{\mu^2-\omega_c^2}\ ,
    \quad\text{where}\quad
    \mu^2>\omega_c^2\ ,
\end{align}
and introducing
\begin{align}
    z\equiv2\epsilon x\ ,
\end{align}
the radial equation is shown to have the form of the Whittaker's equation,
\begin{align}
    z^2\frac{\text{d}^2W(z)}{\text{d}z^2}=\left[\frac{z^2}{4}-kz+\left(p^2-\frac{1}{4}\right)\right]W(z)\ ,
\end{align}
where
\begin{align}
    k=\frac{M\omega_c(M\omega_c+eQ_e)}{\epsilon}-\epsilon\ .
\end{align}
Whittaker's equation has a regular singular point at $z=0$ and an irregular singular point at $z=\infty$, thus being of the confluent hypergeometric type. The bounded solutions of Whittaker's equation have the form
\begin{align}
    W(z)=z^{p+1/2}e^{-z/2}\sum_{j=0}^\infty a_{(j)}z^j\ .
\end{align}
The series must be finite, which is achieved by the quantization condition
\begin{align}
    k=\frac{1}{2}+p+n\ ,
    \quad
    n\in\mathbb{N}_0\ .
    \label{eq:quant}
\end{align}
The regularity of the radial function requires $p\geq-1/2$, which implies that $k>0$. Fixing $\{Q_e\mu,Q_m\mu\}$, Eq.~\eqref{eq:quant} is a quantization condition on $M\mu$, the mass of the extremal KN BH. It reduces to the condition reported by Hod for $Q_e=Q_m=0$~\cite{Hod:2012px}. Solutions to Eq.~\eqref{eq:quant} are thus referred to \textit{Hod points}. Setting either the electric or the magnetic charge to zero and fixing the other one, Eq.~\eqref{eq:quant} has two solutions: the solution with the greatest (lowest) value of $a\mu$ is said to be of Kerr-(Reissner-Nordstr\"om-)type. The following tables show low-$n$ Hod points for values of $\{Q_m\mu,q,\ell,m\}$ considered throughout Section~\ref{section2}.

\begin{table}[h]
    \centering
    \begin{tabular}{c|cc}
        \toprule
        \toprule
        $n$ & $M\mu \approx a\mu$ & $\omega/\mu$ \\
        \hline
        $0$ & $0.255137$ & $0.979867$ \\
        $1$ & $0.251605$ & $0.993621$ \\
        $2$ & $0.250765$ & $0.996950$ \\
        $3$ & $0.250445$ & $0.998224$ \\
        $4$ & $0.250290$ & $0.998840$ \\
        \bottomrule
        \bottomrule
    \end{tabular}
    \label{tab:2}
    \caption{Kerr-type Hod points for $Q_e\mu=0.0$, $Q_m\mu=10^{-18}$, $|q|=\ell=m=1/2$.}
\end{table}

%\begin{table}[h]
%    \centering
%    \begin{tabular}{c|ccc|ccc}
%        \toprule
%        \toprule
%        & \multicolumn{3}{c|}{Reissner-Nordstr\"om-type} & \multicolumn{3}{c}{Kerr-type} \\
%        \hline
%        $n$ & $M\mu$ & $a\mu$ & $\omega/\mu$ & $M\mu$ & $a\mu$ & $\omega/\mu$ \\
%        \hline
%        $0$ & $0.102359$ & $0.021847$ & $0.997142$ & $0.255136$ & $0.254940$ & $0.979868$ \\
%        $1$ & $0.102369$ & $0.021897$ & $0.999049$ & $0.251605$ & $0.251406$ & $0.993621$ \\
%        $2$ & $0.102372$ & $0.021910$ & $0.999532$ & $0.250765$ & $0.250565$ & $0.996950$ \\
%        $3$ & $0.102373$ & $0.021915$ & $0.999722$ & $0.250445$ & $0.250245$ & $0.998224$ \\
%        $4$ & $0.102374$ & $0.021918$ & $0.998816$ & $0.250290$ & $0.250090$ & $0.998840$ \\
%        \bottomrule
%        \bottomrule
%    \end{tabular}
%    \caption{$Q_m\mu=10^{-2}$, $\ell=m=|q|=1/2$.}
%\end{table}

\begin{table}[h]
    \centering
    \begin{tabular}{c|ccc|ccc}
        \toprule
        \toprule
        & \multicolumn{3}{c|}{Reissner-Nordstr\"om-type} & \multicolumn{3}{c}{Kerr-type} \\
        \hline
        $n$ & $M\mu$ & $a\mu$ & $\omega/\mu$ & $M\mu$ & $a\mu$ & $\omega/\mu$ \\
        \hline
        $0$ & $0.102359$ & $0.021847$ & $0.997142$ & $0.254171$ & $0.233672$ & $0.980124$ \\
        $1$ & $0.102369$ & $0.021897$ & $0.999049$ & $0.250645$ & $0.229832$ & $0.993692$ \\
        $2$ & $0.102372$ & $0.021910$ & $0.999532$ & $0.249804$ & $0.228915$ & $0.996981$ \\
        $3$ & $0.102373$ & $0.021915$ & $0.999722$ & $0.249483$ & $0.228565$ & $0.998241$ \\
        $4$ & $0.102374$ & $0.021918$ & $0.998816$ & $0.249328$ & $0.228396$ & $0.998851$ \\
        \bottomrule
        \bottomrule
    \end{tabular}
    \label{tab:3}
    \caption{Hod points for $Q_e\mu=0.0$, $Q_m\mu=0.10$, $|q|=\ell=m=1/2$.}
\end{table}

\begin{table}[h]
    \centering
    \begin{tabular}{c|ccc|ccc}
        \toprule
        \toprule
        & \multicolumn{3}{c|}{Reissner-Nordstr\"om-type} & \multicolumn{3}{c}{Kerr-type} \\
        \hline
        $n$ & $M\mu$ & $a\mu$ & $\omega/\mu$ & $M\mu$ & $a\mu$ & $\omega/\mu$ \\
        \hline
        $0$ & $0.260113$ & $0.071823$ & $0.986348$ & $0.534882$ & $0.472862$ & $0.927731$ \\
        $1$ & $0.260351$ & $0.072682$ & $0.994750$ & $0.508572$ & $0.442883$ & $0.973817$ \\
        $2$ & $0.260424$ & $0.072941$ & $0.997272$ & $0.501289$ & $0.434500$ & $0.987319$ \\
        $3$ & $0.260454$ & $0.073051$ & $0.998338$ & $0.498467$ & $0.431241$ & $0.992642$ \\
        $4$ & $0.260470$ & $0.073108$ & $0.998884$ & $0.497108$ & $0.429670$ & $0.995222$ \\
        \bottomrule
        \bottomrule
    \end{tabular}
    \label{tab:4}
    \caption{Hod points for $Q_e\mu=0.0$, $Q_m\mu=0.25$, $|q|=\ell=m=1$.}
\end{table}

%\begin{table}[h]
%    \centering
%    \begin{tabular}{ccc}
%        \toprule
%        \toprule
%        $\ell=m$ & $M\mu\approx a\mu$ & $\omega/\mu$ \\
%        \midrule
%        $1/2$ & $0.255137$ & $0.979867$ \\
%        $1$   & $0.540335$ & $0.925352$ \\
%        $3/2$ & $0.2299$ & $0.9933$ \\
%        \bottomrule
%        \bottomrule
%    \end{tabular}
%    \caption{$\ell=m=-q$, $n=0$, $Q_m\mu=10^{-1}$.}
%\end{table}

\newpage

%%%%%%%%%%%%%%%%%%%%%%%%%%%%%%%%%%%%%%%%%%%%%%
\bibliographystyle{jhep}  
\bibliography{references}

\end{document}